%% file: paper.tex
\documentclass[10pt,journal,letterpaper]{IEEEtran}
\IEEEoverridecommandlockouts 

\usepackage{paper}
\usepackage{mathdefs}

\begin{document}

\title{Utility Maximization for Single-Station User Association in Downlink Cellular Networks}

\author{%
  Jeffrey~Wildman,~\IEEEmembership{Member,~IEEE,} 
and Steven~Weber,~\IEEEmembership{Senior~Member,~IEEE}%
  \thanks{%
    J. Wildman and S. Weber are with the Department of Electrical and Computer Engineering, Drexel University, Philadelphia, PA, USA (email: jeffrey.wildman@gmail.com and sweber@coe.drexel.edu).  The contact author is S. Weber.%
  }%
  \thanks{%
    An earlier version of this work was presented at the 2015 Allerton Conference on Communications, Control, and Computing (Allerton) \cite{WilOsmWeb2015-ALLERTON}.  This work is partially supported by the National Science Foundation, Award \#CNS-1457306.%
  }
}

\maketitle

\begin{abstract}
We study network utility maximization (NUM) in the context of cellular single station association (SSA) policies, which assigns each mobile user (MU) to a single base station (BS).  We measure an SSA policy in terms of the induced \(\alpha\)-proportional fairness utility of each user's downlink rate, summed over all users. The general SSA NUM problem involves choosing an optimal {\em association} from MUs to BSs as well as an optimal {\em allocation} of BS resources to associated MUs. Finding an exact solution to  such centralized user association problems is well-known to be NP-hard.  Our contributions are as follows: $i)$ we give an explicit solution for the optimal BS allocation for a given SSA, which establishes SSA NUM as a purely combinatiorial problem; $ii)$ we establish the integrality gap for the association problem to be one, and prove the relaxation to be a non-convex optimization problem; $iii)$ we provide both centralized and distributed greedy algorithms for SSA, both with and without the exchange of instantaneous rate information between users and stations.  Our numerical results illustrate performance gains of three classes of solutions: $i)$ SSA solutions obtained by greedy rounding of multi-station associations (a centralized convex program), $ii)$ our centralized and distributed greedy algorithms with/without rate information exchanged, and $iii)$ simple association heuristics.  
\end{abstract}

\begin{IEEEkeywords}
  Cellular network, downlink, user association, resource allocation.
\end{IEEEkeywords}

\section{Introduction}\label{sec:introduction}

\input{introduction}

\section{MSA Problem Statement}\label{sec:problem-statement}

\input{problem-statement}

\section{SSA w/ Optimal Resource Allocation}\label{sec:num-ssa-optalloc}

\input{num-ssa-optalloc}

\section{SSA w/ Uniform Resource Allocation}\label{sec:num-ssa-unialloc}

\input{num-ssa-unialloc}

\section{Algorithms for General \(\alpha\)}\label{sec:num-ssa-algos}

\input{num-ssa-algos}

\section{Numerical Results}\label{sec:results}

\input{results}
\section{Conclusion}\label{sec:conclusions}

We study network utility maximization (NUM) within the context of multi-station association (MSA) and single-station-association (SSA) in cellular networks.  We separate out the association decision of each MU from the resource allocation of each BS, and highlight both optimal and uniform allocation decisions.  
We establish the integrality gap and non-convexity of SSA NUM programs over regimes of \(\alpha\)-utility fairness measures, for both optimal and uniform resource allocations.
Specifically, we show there is an integrality gap of \(1\) for \(\alpha \in [0,2]\), \ie, integer solutions suffice to solve the SUA NUM relaxation.
Interestingly, the convex MSA NUM problem provides the basis for a natural rounding algorithm to feasible SSA solutions that outperform other greedy algorithm approaches proposed in this paper.  Our numerical investigations identify the various performance costs incurred in $i)$ requiring SSA solutions vs.\ allowing MSA solutions, $ii)$ using greedy algorithms vs.\ rounding MSA solutions, $iii)$ using distributed vs.\ centralized greedy algorithms, $iv)$ exchanging vs.\ not exchanging instantaneous rate information, $v)$ using greedy algorithms vs.\ using simple association heuristics, and $vi)$ using optimal vs.\ uniform resource allocations at the BS. 

Natural extensions to this work include $i)$ approximation ratio guarantees for optimal MSA vs.\ optimal SSA solutions, $ii)$ approximation ratio guarantees for the sum utility of an SSA with optimal vs.\ uniform resource allocation, and $iii)$ dynamic algorithms for updating an SSA solution in the presence of changes in either the set of MUs or the set of BSs.  

\appendix

\input{appendix}

\section*{Acknowledgment}

The authors gratefully acknowledge the contributions of Ali Shokoufandeh and Yusuf Osmanl{\i}o\u{g}lu from the Drexel University Department of Computer Science on earlier versions of this work \cite{WilOsmWeb2015-ICC,WilOsmWeb2015-ALLERTON}.


\bibliographystyle{IEEEtran}
\bibliography{IEEEabrv,refs}



\end{document}

%% file: introduction.tex
\subsection{Motivation}\label{sec:motivation}

\IEEEPARstart{T}he transition from traditional cellular networks to heterogeneous networks (HetNets) has opened up many research and design questions, including user association; these are gathered and detailed by Andrews~\cite{And2013} and Ghosh \ea~\cite{GhoManRat2012}.
Finding an exact solution to many centralized user association problems is known to be NP-hard \cite{LuoZha2008}, and so we are motivated to explore the performance of simple heuristics for such problems.

We focus in this paper on single station association (SSA) policies where each mobile user (MU) is permitted to associate with exactly one base station (BS); the associated downlink rates seen by each user are assigned an $\alpha$-fair utility, and the objective is to identify associations that maximize the sum user utility.  This problem falls under the well-studied network utility maximization (NUM) framework.  Central to our study is an investigation of the impact of the $\alpha$ (fairness) parameter on the structure of the associated optimization problem and its solution.

In our previous work \cite{WilOsmWeb2015-ICC,WilOsmWeb2015-ALLERTON} (joint with Osmanl{\i}o\u{g}lu and Shokoufandeh), we considered the minimum delay (\(\alpha=2\)) SSA problem, observed that the natural relaxation is a nonconvex quadratic optimization program, and explored the integer linearization of quadratic reformulations of the core problem \cite{WilOsmWeb2015-ICC}.
Subsequently, we improved our integer linearization by applying it directly to the min-delay SSA problem and leveraged primal-dual algorithms for its approximation \cite{WilOsmWeb2015-ALLERTON}.  This paper extends our previous work by $i)$ establishing {\em optimal} allocations for each BS across the associated MUs for a given association, and $ii)$ providing centralized and distributed greedy algorithms for finding associations.

\subsection{Related Work}\label{sec:related-work}


The objective of user association is typically maximization of MU rates.
The core problem often involves combinatorial optimization by mapping MUs to BSs \cite{SanWanMad2008, YeRonChe2013, SheYu2014}.
An alternative approach to cell association is the use of stochastic geometry to characterize the outage probability of distribution of rates of a typical MU in the network \cite{JoSanXia2012, SinDhiAnd2013, LinYu2013}.  This paper studies SSA as a combinatorial optimization problem and does not incorporate stochastic geometry.

{\em MSA.} Recently there has been interest in user associations that permit multi-station association (MSA), \ie, each MU simultaneously associates with and receives information from multiple BSs.
Interest in optimal MSA is motivated by recent developments of coordinated multi-point (CoMP) for the LTE Advanced standard, under which multiple BSs coordinate their transmissions to an MU to either provide a diversity or multiplexing gain (or both) \cite{SawKisMor2010,LeeKimLee2012,LeeSeoCle2012,TaoNagTak2012}.
Bejerano \ea~\cite{BejHanLi2007} permit multi-cell association in approximation algorithms for max-min fair optimization of MU rates.  Because the set of MSAs contains the set of SSAs, the optimal sum-user utility of an MSA solution is an upper bound on that of an SSA solution.

{\em SSA.} Our primary focus in this paper is on the SSA problem. 
The closest reference to our approach is Ye \ea~\cite{YeRonChe2013}, which addresses the SSA problem in the special case of \(\log\)-utility ($\alpha=1$), \ie, proportional fairness.  The authors relax the combinatorial SSA problem and obtain a primal-dual algorithm and prove it converges to the solution of the relaxation. In the present paper, we also consider an SSA NUM relaxation, but for a {\em general} \(\alpha\)-proportional fairness measure.  One of our key results is that integer solutions suffice to solve the relaxed problem when \(\alpha \in [0,2]\).
We emphasize, however, that  integer solution optimality for a relaxed SSA NUM problem does not imply the same for the corresponding MSA NUM problem.

{\em Fairness.} Fairness of association schemes has been addressed by several works \cite{SanWanMad2008, SonChoDeV2009, BejHanLi2007, KimDeVYan2012}.
Sang \ea~\cite{SanWanMad2008} propose a cross-layer algorithm to maximize the network's sum, weighted, \(\alpha\)-proportional fair utility.
Son \ea~\cite{SonChoDeV2009} propose off- and on-line algorithms to compute handoff and association rules to achieve network-wide proportial rate fairness across MUs with \(\log\)-based utility.
Bejerano \ea~\cite{BejHanLi2007} and Sun \ea~\cite{SunHonLuo2015} explore association rules that promote max-min rate fairness across MUs.
Kim \ea~\cite{KimDeVYan2012} propose a \(\alpha\)-optimal user association rule related to \(\alpha\)-proportional fairness \cite{MoWal2000} that provides a tradeoff between individual user rate maximization and load balancing across BSs.
Our results are complimentary to the above references in that we study $i)$ MU association vs.\ BS resource allocation, $ii)$ the integrality gap of the relaxation of the SSA problem and its non-convexity, and $iii)$ the performance of simple greedy algorithms to obtain SSAs; these results are not found in the above work.

\subsection{Outline and Contributions}\label{sec:contributions}

\tabref{notation} lists the mathematical notation used in the paper.  In \secref{problem-statement} we introduce a general MSA NUM problem parameterized by an \(\alpha\)-utility fairness measure, establish that the MSA NUM problem is convex (\prpref{num-convex}), and present simulation results (\figref{MSA}) that suggest the empirical observation that the solutions of the MSA NUM problem are often ``SSA-like'', meaning most MUs associate with a single BS even when not constrained to do so.  This suggests that it should be possible to obtain ``good'' SSA solutions by greedy rounding of MSA solutions; we investigate this approach numerically in \secref{results}.

In \secref{num-ssa-optalloc}, we specialize the MSA NUM to the case of SSA associations/allocations and study the structural properties of the problem under optimal BS resource allocations.  Given an association, $i)$ we establish the resulting optimization problem for each BS to allocate resources across its associated MUs is convex (\prpref{num-ssa-convex}), and $ii)$ admits an explicit solution (\prpref{num-ssa-solution}), wherein $iii)$ the sensitivity of the optimal allocation to the instantaneous rate qualitatively depends upon $\alpha \gtrless 1$ (\prpref{num-ssa-sensitivity}).  We then consider the optimization problem obtained by integer relaxation of the SSA NUM problem and show the integrality gap of the relaxation to be unity (\thmref{num-ssa-optalloc-int-gap}) for $\alpha \in [0,2]$, which means the set of solutions of the relaxation must contain an integral (SSA) solution.  Unfortunately, but unsurprisingly, the relaxation is a nonconvex optimization problem for all $\alpha \neq 1$ (\prpref{num-rssa-optalloc-convexity}).  It is natural to inquire whether the relaxation is convexifiable, \eg, through geometric programming (GP); we establish a ``standard'' GP approach essentially fails for this problem in \secref{ssa-gp}.

In \secref{num-ssa-unialloc}, we restrict our attention to BS resource allocations that {\em uniformly} share BS resources among associated users, \ie, the BS ignores the various instantaneous rates achievable between the BS and each associated MU, in contrast to the {\em optimal} allocations discussed in \secref{num-ssa-optalloc}.  The motivation behind this study is the scenario where the BS may not have access to the instantaneous rate information between it and its associated MUs, which are necesssary to compute the optimal resource allocation.  Similar to the optimal case, we establish the integrality gap of the relaxation of the association problem with uniform BS resource allocation is also one for $\alpha \in [0,2]$ (\thmref{num-ssa-unialloc-int-gap}), but again, the relaxation is a nonconvex program for all $\alpha\neq 1$ (\prpref{num-rssa-unialloc-convexity}).

In \secref{num-ssa-algos}, we present three greedy algorithms for producing feasible SSA assocations; the three algorithms are differentiated by the amount of instantaneous downlink rate information they employ.  A fourth algorithm uses greedy rounding to obtain an SSA solution from a solution of an MSA problem.  The performance of these four algorithms is compared in \secref{results}, where our performance metrics are $i)$ the absolute and relative loss of sum utility of the obtained SSA solutions relative to the optimal MSA solution, $ii)$ the Chiu-Jain fairness measure of the associated downlink user rates, and $iii)$ the sum user network downlink throughput, all four metrics are swept over the $\alpha$ utility parameter.  Simulation results in \figref{secV} show an ordering of the proposed solutions along with the performance of  simple heuristics.  Additional results in \figref{secV2} show the relative importance of optimal allocations vs.\ uniform allocations.  

In \secref{conclusions}, we conclude our work and touch upon ideas for future investigation.
Finally, for clarity, long proofs are presented in the Appendix.

The major contributions of this paper are as follows.
\begin{enumerate}
\item We separate the association decision of each MU to select a single BS from the allocation decision of how each BS shares resources across its associated MUs.  Whereas the association problem is a combinatorial optimization, the allocation problem is a continuous optimization.  We establish the optimal allocation for a given association, thereby transforming the optimal SSA NUM problem to one that is purely combinatorial.
\item We establish that the integrality gap of the SSA NUM problem is one relative to its natural relaxation for all $\alpha \in [0,2]$, which means the solution set of the relaxation will contain at least one integral solution to the original combinatorial problem.  Unfortunately, for all $\alpha \neq 1$, the relaxation is non-convex, so the essential difficulty of the SSA NUM problem is in solving a nonconvex optimization problem. 
\item We provide simple greedy algorithms, both centralized and distributed, including variants that both assume and do not assume instantaneous rate information exchange, for obtaining (in general, suboptimal) associations.  Our numerical investigations identify the various performance costs incurred in $i)$ requiring SSA solutions vs.\ allowing MSA solutions, $ii)$ using greedy algorithms vs.\ rounding MSA solutions, $iii)$ using distributed vs.\ centralized greedy algorithms, $iv)$ exchanging vs.\ not exchanging instantaneous rate information, $v)$ using greedy algorithms vs.\ using simple association heuristics, and $vi)$ using optimal vs.\ uniform resource allocations at the BS.  
\end{enumerate}

\begin{table}
  \caption{Notation}%
  \label{tab:notation}%
  \centering%
  \renewcommand{\arraystretch}{1.3}%
  \begin{tabular}{llr} \hline \hline
    \secref{problem-statement} & MSA Problem Statement & \\ \hline
    $\mathsf{sinr}_{ub}$ & $b \to u$ instantaneous sinr & \eqref{sinrdef} \\
    $r_{ub}$ & $b \to u$ instantaneous downlink rate & \eqref{instantaneous-rate} \\
    $y_{ub}$ & $b \to u$ resource allocation & \\
    $R_u(\ybf)$ & downlink rate to $u$ under allocation $\ybf$ & \eqref{sum-rate} \\
    $\Ymc$ & set of feasible BS allocations & \eqref{allocation-space} \\
    $U_{\alpha}(R)$ & $\alpha$-proportional fair utility for DL rate $R$ & \eqref{utility} \\
    $\fMSA$ & optimal sum utility over all MSAs & \eqref{num} \\
    $\Jmc_u$ & Chiu-Jain fairness of DL rates from BSs for $u$ & \eqref{cjfu} \\ \hline \hline
    \secref{num-ssa-optalloc} & SSA w/ Optimal Resource Allocation & \\ \hline
    $z_{ub}$ & indicator that MU $u$ associates (only) with BS $b$ \\
    $\Zmc$ & set of feasible SSA associations & \eqref{zmcdef} \\
    $R_{u}(\ybf, \zbf)$ & DL rate for MU $u$ under alloc.\ $\ybf$ and assoc.\ $\zbf$ & \eqref{ssa-sum-rate} \\
    $\fSSA,\fSO$ & opt.\ sum utility over all SSAs, w/ opt.\ BS alloc.\ & \eqref{num-ssa} \\
    $\fSSA(\zbf,b)$ & opt.\ sum utility at BS $b$ under assoc.\ $\zbf$ & \eqref{num-ssa-decomp} \\
    $y_{ub}^*(\zbf)$ & opt.\ allocation $y_{ub}$ for association $\zbf$ & \eqref{num-ssa-solution} \\
    $\Xmc$ & integer relaxation of feasible SSA set $\Zmc$ & \eqref{xmcdef} \\
    $\fRSO$ & opt.\ sum utility over $\Xmc$, w/ opt.\ BS alloc.\ & \\
    $\fSSA(\epsilon)$ & $\epsilon$-SSA NUM problem & \eqref{num-ssa-gp} \\ \hline \hline
    \secref{num-ssa-unialloc} & SSA w/ Uniform Resource Allocation & \\ \hline
    $\fSU$ & opt.\ sum utility over all SSAs, w/ unif.\ BS alloc.\ & \\
    $\fRSU$ & opt.\ sum utility over $\Xmc$, w/ unif.\ BS alloc.\ & \\ \hline \hline
  \end{tabular}
\end{table}

%% file: problem-statement.tex
In this section, we present a network utility maximization (NUM) problem called multi-station assignment (MSA), parameterized by an \(\alpha\)-proportional fair utility measure.  The distinguishing feature of the MSA NUM problem is that each MU is allowed to associate with multiple BSs, and the perceived utility by each MU is the overall downlink rate obtained by summing the allocations from each BS to that MU, with each allocation weighted by the corresponding instantaneous rate.
In later sections, we will specialize this problem formulation to the case of single-station-association (SSA), denoted \(\fSSA\), where each MU associates with exactly one BS.
For convenience, we often write only the index for the following index-support sets: \(u \in \Umc\), \(v \in \Umc\), \(a \in \Bmc\), \(b \in \Bmc\), \(\xbf \in \Xmc\), \(\ybf \in \Ymc\), and \(\zbf \in \Zmc\).

Consider a set of BSs \(\Bmc\) at locations \(\{l_b, \forall b \in \Bmc\}\) and a set of MUs \(\Umc\) at locations \(\{l_u, \forall u \in \Umc\}\), where both sets exist within a bounded arena \(\Amc \subset \Rbb^2\).
We model the {\em instantaneous downlink rate} from BS \(b\) to MU \(u\) using the Shannon rate of their point-to-point channel, treating interference from other BSs as additive white Gaussian noise:
\begin{equation}\label{eq:instantaneous-rate}
  r_{ub} = \log(1 + \mathsf{sinr}_{ub}),
\end{equation}
where the signal-to-interference-plus-noise ratio (SINR) is:
\begin{equation}\label{eq:sinrdef}
  \mathsf{sinr}_{ub} = \frac{p_b g(l_b,l_u)}{\sum_{a\in\Bmc \setminus b} p_a g(l_a,l_u) + N},
\end{equation}
as determined by the BS transmission powers \(\{p_b, \forall b \in \Bmc \}\), the channel attenuation function \(g(l,l') = \|l - l'\|_2^{-\gamma}\) with pathloss exponent \(\gamma \geq 2\), and background noise power \(N \geq 0\).  

We assume each BS multiplexes its resources (\eg, time or frequency) across its associated users.
Let \(y_{ub}\) represent the fraction of resources from BS \(b\) allocated to MU \(u\), so that the sum downlink rate to MU \(u\) is given by:
\begin{equation}\label{eq:sum-rate}
  R_{u}(\ybf) = \sum_{b} r_{ub}y_{ub},
\end{equation}
for $\ybf = (y_{ub}, u \in \Umc, b \in \Bmc)$ the $|\Umc| \times |\Bmc|$ resource allocation matrix.  
We will require that each BS be constrained to a fixed set of resources that may be allocated, which we set to unity at each BS, \ie, $\sum_u y_{ub} = 1$ for each $b \in \Bmc$.
The resulting set of feasible BS allocations is denoted \(\Ymc\):
\begin{equation}\label{eq:allocation-space}
  \textstyle \Ymc \!=\! \left\{\ybf: \ybf \in \Rbb_{+}^{|\Umc| \times |\Bmc|}, \sum_{u} y_{ub} = 1, \forall b\in\Bmc\right\}.
\end{equation}
Under this scenario, each MU may associate with (receive non-zero allocations from) multiple stations -- we denote this scenario as multi-station association (MSA).
Finally, we will consider a diverse family of NUM problems derived from \(\alpha\)-proportional fairness measures \cite{MoWal2000,UchKur2011}, defined as:
\begin{equation}\label{eq:utility}
  \utility{R} = \begin{cases}
    \frac{R^{1-\alpha}}{1-\alpha}, & \alpha \geq 0, \alpha \neq 1 \\
    \log(R), & \alpha = 1
  \end{cases}.
\end{equation}

Under MSA, the objective is to find an optimal allocation of resources \(\ybf\) that maximizes the sum utility of the sum downlink rates at each of the MUs.
The resulting MSA NUM problem and solution, denoted \(\fMSA\), is as follows:
\begin{equation}
  \fMSA \equiv \max_{\ybf \in \Ymc} \sum_{u} \utility{R_{u}(\ybf)} \label{eq:num}.
\end{equation}

\begin{proposition}[Convexity of \(\fMSA\)]\label{prp:num-convex}
  The MSA NUM problem, \(\fMSA\), given by \eqref{num} is convex.
\end{proposition}
\begin{IEEEproof}
  We follow the characterization in \cite[\S4.2.1]{BoyVan2004}.
  The objective function is a non-negative weighted sum of concave functions composed with an affine function over \(\ybf\), ultimately yielding a concave function to be maximized.
  The equality constraints embedded in \(\Ymc\) are affine and the domain \(\ybf \in \Rbb_{+}^{|\Umc|\times|\Bmc|}\) is convex.
\end{IEEEproof}

While the MSA NUM is convex and its centralized solution may be obtained without much difficulty, the resulting solutions may require MUs to associate with and receive resource allocations from multiple BSs.
MSA solutions are supported by upcoming features of cellular systems (CoMP), but not all users may be equipped with the necessary hardware and/or software.
This constraint motivates the study of single-station-association (SSA) formulations, where each MU may receive resource allocations from at most one BS.

\begin{figure}[t!]
  \centering%
  \includegraphics[width=\columnwidth]{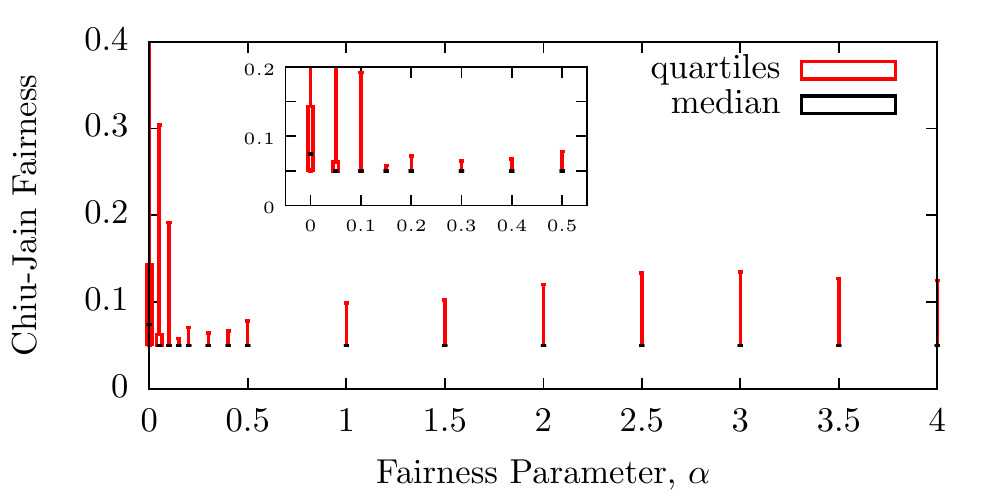}%
  \caption{%
    The \(\{0,25,50,75,100\}\)\% percentiles of the Chiu-Jain fairness measure of each MU's rate vector \((r_{ub}y_{ub}^*, b\in\Bmc)\) under the optimal MSA allocation.
    The \(\{0,25,75,100\}\)\% percentiles are plotted using a red box and whiskers, while the \(50\%\) percentile (median)  is plotted in black.
    The percentiles are averaged over \(100\) independent samples where each network sample consists of \(100\) MUs and \(20\) BSs, with positions generated uniformly at random within a square \(1000 \times 1000\) meter arena.
    Each BS transmits at \(1000\) mW with a signal bandwidth of \(1.2\) MHz, background noise is assumed to be \(-90\) dBm, and the pathloss constant is set to \(3\).
  }%
  \label{fig:MSA}%
\end{figure}

Additionally, we refer to \figref{MSA}, which demonstrates the degree to which optimal MSA allocations \(\ybf^*\) display qualities that are ``SSA-like'', at least for the (reasonable) scenario where BSs and MUs are distributed uniformly at random over the network arena.   
Plotted are percentiles of the Chiu-Jain fairness measure \cite{JaiChiHaw1984} over each MU's rate vector:
\begin{equation}
  \Jmc_u([r_{ub}y_{ub}^*, \forall b\in\Bmc]) = \frac{\left(\sum_{b} r_{ub}y_{ub}^*\right)^2}{|\Bmc| \sum_{b} \left(r_{ub}y_{ub}^*\right)^2}, \forall u \in \Umc. \label{eq:cjfu}
\end{equation}
The Chiu-Jain fairness measure used in this way ranges over the interval \([1/\Bmc,1]\); the endpoints of this interval capture $i)$ the scenario where an MU receives a non-zero rate from exactly one BS (the least equitable, \(\Jmc_u = 1/\Bmc\)), and $ii)$ the scenario in which an MU receives equal rates from all BSs (the most equitable, \(\Jmc_u = 1\)).
With the exception of small \(\alpha\) (less than $0.05$), we observe that over \(75\%\) of MUs have rate vectors that are SSA-like, \ie, the sum rate \eqref{sum-rate} of those MUs is comes from a single BS.
However, the \(100\%\) percentile shows that at least one MU may receive significant downlink rates from more than one BS.
These results highlight the possibility that SSA solutions may represent a sufficient subspace over which to optimize network utility.

%% file: num-ssa-optalloc.tex
In this section, we specialize the MSA NUM problem formulation to the case of single-station-association (SSA), denoted \(\fSSA\), where each mobile user (MU) associates with exactly one base station (BS).
We provide structural analysis of two SSA NUM problem formulations where BSs are free to optimize resource allocation.
We first capture the SSA constraint with the addition of association variables \(\zbf\), and note that the SSA problem is a mixed-integer nonlinear program.
Under a fixed association scheme, we find that the problem may be decomposed into smaller, convex programs in the remaining BS resource allocation variables with well-defined solutions.
Later in this section, we detail an alternative attempt to enforce SSA solutions by introducing pairwise allocation constraints on the allocation variables.
We note that the problem \(\fSSA\) becomes non-convex, but may be relaxed to that of a Geometric Program (GP).
While convexifiable, the GP is shown to produce solutions that do not faithfully capture the desired SSA constraints.

\subsection{SSA Formulation}

Under SSA, each MU may associate with, or receive allocations from, exactly one BS.
SSA allocations may be enforced by introducing a set of association variables \(\zbf = (z_{ub}, u \in \Umc, b \in \Bmc)\), where \(z_{ub}=1\) iff MU \(u\) is assigned to BS \(b\).
The resulting set of feasible associations is denoted \(\Zmc\):
\begin{equation}
  \textstyle \Zmc \!=\! \left\{\zbf: \zbf \in \Zbb_{+}^{|\Umc| \times |\Bmc|}, \sum_{b} z_{ub} = 1, \forall u\in\Umc\right\}, \label{eq:zmcdef}
\end{equation}
and the sum downlink rate to MU \(u\) may be expressed as:
\begin{equation}\label{eq:ssa-sum-rate}
  R_{u}(\ybf, \zbf) = \sum_{b} r_{ub}y_{ub}z_{ub}.
\end{equation}

The SSA NUM problem is now stated:
\begin{equation}\label{eq:num-ssa}
  \fSSA \equiv \max_{\substack{\ybf \in \Ymc,\\ \zbf \in \Zmc}} \sum_{u} \utility{R_{u}(\ybf,\zbf)}.
\end{equation}
Note, under a feasible SSA allocation \(\ybf\), the summation over \(b\) in \eqref{ssa-sum-rate} contains exactly one positive term and may be equivalently written as:
\begin{equation}\label{eq:objfun-manip}
  \sum_{u} \utility{R_{u}(\ybf,\zbf)} = \sum_{ub} \utility{r_{ub}y_{ub}}z_{ub}.
\end{equation}


We note that for a fixed set of association variables \(\zbf\), we may partition the set of MUs by their associated BS.
Let \(\Umc_{b}\) be the set of MUs associated with BS \(b\): \(\Umc_{b} \equiv \{u | z_{ub} = 1\}\).
Under any fixed association \(\zbf\), the problem \(\fSSA\) decomposes into smaller, convex problems at each BS \(b\):
\begin{align}
  \fSSA(\zbf,b) \equiv \max & \sum_{u\in\Umc_{b}} U_{\alpha}(r_{ub}y_{ub})    \label{eq:num-ssa-decomp} \\
  \textrm{s.t.}           & \sum_{u\in\Umc_{b}} y_{ub} = 1                  \label{eq:num-ssa-decomp-const1} \\
                          & y_{ub} \in \Rbb_{+}, \quad \forall u\in\Umc_{b} \label{eq:num-ssa-decomp-const2}.
\end{align}

\begin{proposition}[Convexity of \(\fSSA(\zbf,b)\)]\label{prp:num-ssa-convex}
  Under a fixed association \(\zbf\), the SSA NUM subproblems, \(\{\fSSA(\zbf,b), \forall b\in\Bmc\}\), given by \eqref{num-ssa-decomp} are convex.
\end{proposition}
\begin{IEEEproof}
  Upon decomposing \(\fSSA\) into subproblems, convexity follows similarly using the proof of \prpref{num-convex}.
\end{IEEEproof}

\begin{proposition}[Solution of \(\fSSA(\zbf,b)\)]\label{prp:num-ssa-solution}
  Let \(\zbf\) be a fixed association and \(b\in\Bmc\) be a fixed BS.
  When \(\alpha = 0\), solutions to \(\fSSA(\zbf, b)\) are characterized by all \(\ybf \in \Ymc\) that only allocate BS resources to associated MUs with the largest instantaneous rate:
  \begin{align}
    y_{ub}^*(\zbf) &= 0, \quad \forall u \notin \left\{\argmax_{v\in\Umc} r_{vb}z_{vb}\right\}.
  \end{align}

  When \(\alpha > 0\), the solution to \(\fSSA(\zbf,b)\) has the following structure:
  \begin{equation}\label{eq:num-ssa-solution}
    y_{ub}^*(\zbf) = \begin{cases}
      \frac{z_{ub}}{1 + \sum_{v}^{v\neq u} z_{vb}}                                  & \alpha = 1 \\
      \frac{s_{ub}z_{ub}}{s_{ub} + \sum_{v}^{v \neq u} s_{vb}z_{vb}}                & \alpha > 0, \alpha \neq 1 \\
      \frac{r_{ub}^{-1}z_{ub}}{r_{ub}^{-1} + \sum_{v}^{v\neq u} r_{vb}^{-1}z_{vb} } & \alpha \to \infty
    \end{cases},
  \end{equation}
  where \(s_{ub} = r_{ub}^{(1-\alpha)/\alpha}\).
\end{proposition}
\begin{IEEEproof}
  See proof in \prfref{num-ssa-solution}.
\end{IEEEproof}


\begin{remark}\label{rem:num-ssa-solution}
  We note that \prpref{num-ssa-solution} is the natural extension of the result by Ye \ea~\cite[Prp.~1]{YeRonChe2013} to the other \(\alpha\)-utility measures.
  When \(\alpha = 1\), we refer to \cite{YeRonChe2013} who showed that the optimal SSA allocation is uniform across associated MUs.
  When \(\alpha = 0\), the NUM problem is equivalent to throughput maximization and the MU with the largest rate is served exclusively.
  When \(\alpha > 0\), the optimal SSA allocation scales with the relative rates of the MUs associated with each BS.
\end{remark}

Given a fixed association policy \(\zbf\), we may also characterize how the optimal allocation policy \(\ybf^*(\zbf)\) behaves as a function of the instantaneous rates in the network:
\begin{proposition}[Sensitivity of \(y_{ub}^*\) to Instantaneous Rates]\label{prp:num-ssa-sensitivity}
  Let \(\zbf\) be a fixed association and \(b\in\Bmc\) be a fixed BS.
  The sensitivity of the optimal resource allocation \(\ybf^*\) to changing instantaneous rates obeys the following:
  \begin{equation}
    \left. \frac{\partial y_{ub}^*}{\partial r_{ub}} \right|_{z_{ub}=1} \begin{cases}
      > 0, &\quad 0<\alpha<1 \\
      = 0, &\quad \alpha=1 \\
      < 0, &\quad 1<\alpha
    \end{cases}.
  \end{equation}
\end{proposition}
\begin{IEEEproof}
  See proof in \prfref{num-ssa-sensitivity}.
\end{IEEEproof}
From \prpref{num-ssa-sensitivity}, we see that a BS will tend to positively reward a larger rate with a greater allocation when \(\alpha < 1\), but reduce resource allocations to larger rates when \(\alpha > 1\).
When the allocation policy is uniform (\(\alpha=1\)), we observe that the optimal solution is insensitive to instantaneous rates.

Under the optimal allocation scheme, the resulting SSA NUM is completely combinatorial in nature and is given by \prpref{num-ssa-optalloc}.
\begin{proposition}[SSA NUM Problem w/ Optimal Allocation]\label{prp:num-ssa-optalloc}
  The SSA NUM problem \eqref{num-ssa} employing optimal resource allocation is denoted \(\fSO\) (S for SSA, O for optimal) and is given by the second column of \tabref{num-ssa}.
\end{proposition}
\begin{IEEEproof}
  See proof in \prfref{num-ssa-optalloc}.
\end{IEEEproof}

Observe that $\fSO$, defined above, equals $\fSSA$, defined in \eqref{num-ssa}: the $\fSSA$ notation is for contrast with the MSA solution $\fMSA$, defined in \eqref{num}, and $\fSO$ is for contrast with the relaxed SSA solution $\fRSO$, defined below.  

As Ye \ea \cite{YeRonChe2013} state, a brute force approach to the combinatorial problem \eqref{num-ssa} has a complexity of \(\Theta(|\Bmc|^{|\Umc|})\).
Hence, we work towards alternative \(\fSSA\) formulations for specific \(\alpha\)-fairness utility functions that are easier to approximate or estimate for larger problem instances.

\begin{table*}[!t]%
  \caption{SSA NUM Formulations}%
  \label{tab:num-ssa}%
  \centering%
  \renewcommand{\arraystretch}{2}%
  \begin{tabular}{c|c|c}
    \(\alpha\) & \(\fSO\) & \(\fSU\) \\ \hline
    \(\alpha = 0\) & \(\max_{\zbf} \sum_{b} \max_{u} r_{ub}z_{ub}\) &
    \(\max_{\zbf} \sum_{ub} r_{ub}z_{ub}\left(1 + \sum_{v}^{v\neq u} z_{vb}\right)^{-1}\) \\
    \(\alpha > 0, \alpha \neq 1\) &
    \(\max_{\zbf} \frac{1}{1\!-\!\alpha} \sum_{ub} r_{ub}^{1-\alpha} z_{ub} \left(1 + \sum_{v}^{v\neq u} \left(\frac{r_{vb}}{r_{ub}}\right)^{\frac{1-\alpha}{\alpha}} z_{vb} \right)^{\alpha-1}\) &
    \(\max_{\zbf} \frac{1}{1\!-\!\alpha} \sum_{ub} r_{ub}^{1-\alpha} z_{ub} \left(1 + \sum_{v}^{v \neq u} z_{vb} \right)^{\alpha-1}\) \\
    \(\alpha = 1\) &
    \(\max_{\zbf} \sum_{ub} z_{ub} \left( \log\!\left( r_{ub} \right) \!-\! \log\!\left( 1 + \sum_{v}^{v\neq u} z_{vb} \right) \right)\) &
    \(\max_{\zbf} \sum_{ub} z_{ub} \left( \log\!\left( r_{ub} \right) \!-\! \log\!\left( 1 + \sum_{v}^{v\neq u} z_{vb} \right) \right)\) \\
    \(\alpha \rightarrow \infty\) &
    \(\max_{\zbf} \min_{u} \left( \sum_{b} z_{ub} \left( r_{ub}^{-1} + \sum_{v}^{v\neq u} r_{vb}^{-1} z_{vb} \right)^{-1} \right)^{-1}\) &
    \(\max_{\zbf} \min_{u} \left( \sum_{b} r_{ub}z_{ub} \left(1 + \sum_{v}^{v\neq u} z_{vb} \right)^{-1} \right)^{-1}\)
  \end{tabular}%
\end{table*}

\subsection{Integrality Gap \& Convexity}\label{sec:optalloc-integrality}

One approach to solving the combinatorial SSA NUM problem is to consider its integer relaxation, denoted \(\fRSO\) (R for relaxed), obtained from \tabref{num-ssa} by replacing \(\zbf \in \Zmc\) with \(\xbf \in \Xmc\).
\(\Xmc\) is the space of relaxed SSA associations (RSSA):
\begin{equation}
  \textstyle \Xmc \!=\! \left\{\xbf: \xbf \in \Rbb_+^{|\Umc|\times|\Bmc|}, \sum_{b} x_{ub} = 1, \forall u\in\Umc\right\}. \label{eq:xmcdef}
\end{equation}
The relationship between optimal solutions of the integer problem and its relaxation is quantified by its \emph{integrality gap} (adapted from Williamson and Shmoys \cite[Def.~5.13]{WilShm2011}).
\begin{definition}[Integrality Gap]\label{def:int-gap}
  The integrality gap between integer program \(\fSO\) over \(\zbf \in \Zmc\) and its relaxation \(\fRSO\) over \(\xbf \in\Xmc\) is the worst case ratio of the value of an optimal solution to the integer program to the value of an optimal solution to its relaxation: \(\fSO(\zbf^*) / \fRSO(\xbf^*)\).
\end{definition}
We first look to quantify the integrality gap associated with rounding non-integer solutions of \(\fRSO\) for different regimes of \(\alpha\) and follow this with a characterization of the convexity of \(\fRSO\) for different regimes of the fairness parameter \(\alpha\).

\begin{remark}\label{rem:ssa-vs-fua}
  Optimal RSSA associations \(\xbf^*\) for \(\fRSO\) are intended to obtain feasible integer SSA associations \(\zbf(\xbf^*)\) (and thus SSA allocations \(\ybf(\zbf(\xbf^*))\) by \prpref{num-ssa-solution}) to \(\fSO\).  Optimal RSSA associations $\xbf^*$ (for which $\sum_b x_{ub}^* = 1$ for each $u \in \Umc$) are distinct from optimal MSA allocations $\ybf^*$ (for which $\sum_u y_{ub}^* = 1$ for each $b \in \Bmc$).
\end{remark}

Due to the fact that \(\fRSO\) is an integer relaxation of \(\fSO\), we immediately know that \(\fSO \leq \fRSO\), but \thmref{num-ssa-optalloc-int-gap} shows that there is no integrality gap in the SSA NUM problem for \(\alpha\in[0,2]\).
In this regime of \(\alpha\), \(\fRSO\) may be attained via integer solutions that simultaneously attain \(\fSO\).
We summarize our findings on the integrality gap in \tabref{num-ssa-int-convex}.
\begin{theorem}[Integrality Gap of \(\fSO\)]\label{thm:num-ssa-optalloc-int-gap}
  When \(\alpha\in[0,2]\), the integrality gap associated with the SSA NUM problem \(\fSO\) is equal to \(1\):
  \begin{equation}
    \fSO(\zbf^*) / \fRSO(\xbf^*) = 1, \quad \alpha \in [0,2].
  \end{equation}
\end{theorem}
\begin{IEEEproof}
  See proof in \prfref{num-ssa-optalloc-int-gap}.
\end{IEEEproof}
\begin{remark}\label{rem:extra-steps}
  Our use of Jensen's Inequality in the proof of \thmref{num-ssa-optalloc-int-gap} prevents us from extending the argument to \(\alpha > 2\).
  In the regime of \(\alpha \in [0,2]\), \thmref{num-ssa-optalloc-int-gap} implies that the optimizers of \(\fRSO\) must contain one or more integer solutions \(\xbf^* \in \Zmc\).
  However, this does not preclude the existence of non-integral optimizers of \(\fRSO\), \ie, \(\xbf^* \in \Xmc \setminus \Zmc\).
\end{remark}


\begin{table*}[!t]
  \caption{Integrality Gap/Convexity of SSA NUM}
  \label{tab:num-ssa-int-convex}
  \renewcommand{\arraystretch}{1.1}%
  \centering
  \begin{tabular}{c|l|l|l|l}
    \(\alpha\)
      & \(\fSO\) Integrality Gap & \(\fRSO\) Convexity
      & \(\fSU\) Integrality Gap & \(\fRSU\) Convexity \\ \hline
    \(\alpha\in[0,1)\)
      & \(=1\) (\thmref{num-ssa-optalloc-int-gap})            & no (\prpref{num-rssa-optalloc-convexity})
      & \(=1\) (\thmref{num-ssa-unialloc-int-gap})            & no (\prpref{num-rssa-unialloc-convexity}) \\
    \(\alpha=1\)
      & \(=1\) (\thmref{num-ssa-optalloc-int-gap})            & yes (\cite{YeRonChe2013})
      & \(=1\) (\thmref{num-ssa-unialloc-int-gap})            & yes (\cite{YeRonChe2013}) \\
    \(\alpha\in(1,2]\)
      & \(=1\) (\thmref{num-ssa-optalloc-int-gap})            & no (\prpref{num-rssa-optalloc-convexity})
      & \(=1\) (\thmref{num-ssa-unialloc-int-gap})            & no (\prpref{num-rssa-unialloc-convexity}) \\
    \(\alpha > 2\)
      & \(\leq 1\)                                            & no (\prpref{num-rssa-optalloc-convexity})
      & \(\leq 1\)                                            & no (\prpref{num-rssa-unialloc-convexity}) \\
    \(\alpha \rightarrow \infty\)
      & \(\leq 1\)                                            & no (\prpref{num-rssa-optalloc-convexity})
      & \(\leq 1\)                                            & no (\prpref{num-rssa-unialloc-convexity})
  \end{tabular}
\end{table*}

While integer relaxation is a valid approach to solving the SSA NUM problem, we find that the relaxed problem \(\fRSO\) generally results in non-convex problems and organize these results in \tabref{num-ssa-int-convex}.
\begin{proposition}[Non-convexity of \(\fRSO\)]\label{prp:num-rssa-optalloc-convexity}
  \(\fRSO\) is non-convex for \(\alpha \geq 0, \alpha \neq 1\) and is convex for \(\alpha=1\).
\end{proposition}
\begin{IEEEproof}
  See proof in \prfref{num-rssa-optalloc-convexity}.
\end{IEEEproof}

Together, \thmref{num-ssa-optalloc-int-gap}, \prpref{num-rssa-optalloc-convexity}, and \tabref{num-ssa-int-convex} establish the essential difficulty in the SSA NUM problem with optimal allocations is the solution of a non-convex optimization problem.

\subsection{SSA Formulation Attempt via Geometric Programming}
\label{sec:ssa-gp}

Although \prpref{num-rssa-optalloc-convexity} establishes that SSA NUM with optimal resource allocation is non-convex as posed, this does not immediately preclude the possibility of a problem tranformation that is convex.  A standard approach to attempting this type of convexification is through geometric programming (GP).  Towards this end we first observe that the SSA NUM problem may alternatively be introduced by adding an additional set of constraints \eqref{num-ssa1-const1} to the MSA problem \eqref{num}, that restricts each MU to receive allocations from at most one BS:
\begin{align}
  \fSSA \equiv \max_{\ybf \in \Ymc} & \sum_{ub} U_{\alpha}(r_{ub}y_{ub}) \label{eq:num-ssa1} \\
  \textrm{s.t. }                       & y_{ua}y_{ub} = 0, \quad \forall u\in\Umc, a\neq b \in\Bmc,  \label{eq:num-ssa1-const1}
\end{align}
where we again apply \eqref{objfun-manip}.  We note that \(\fSSA\) is not convex.
\begin{proposition}[Non-convexity of \(\fSSA\)]\label{prp:num-ssa1-nonconvex}
  The SSA NUM problem, \(\fSSA\), given by \eqref{num-ssa1} is not convex.
\end{proposition}
\begin{IEEEproof}
  Building off of \prpref{num-convex}, we show the non-convexity of the (convex) set $\Ymc$ restricted by the additional constraints \eqref{num-ssa1-const1}.
  Define the following: \(f(\xbf) = x_1 x_2\) be defined over convex domain \(\xbf \in [0,1]^2\).
  Because \(f\) is quadratic, the Hessian of \(f\) is constant over its domain:
  \begin{equation}
    H = \nabla^2 f(\xbf) = \begin{bmatrix}
      0 & 1 \\ 1 & 0
    \end{bmatrix},
  \end{equation}
  whose eigenvalues are \(\lambda(H) = \{\pm 1\}\).
  It follows that \(H\) is not positive semi-definite over the domain of \(f\) and thus \(f\) is not convex.
\end{IEEEproof}

To combat the non-convexity of \(\fSSA\), we may massage its constraints to form a convexifiable geometric program, which we call the \(\epsilon\)-SSA NUM problem, \(\fSSA(\epsilon)\):
\begin{align}
  \fSSA(\epsilon) \equiv \min_{\ybf}  & \frac{1}{\alpha-1} \sum_{ub} (r_{ub} y_{ub})^{1-\alpha} \label{eq:num-ssa-gp} \\
  \textrm{s.t. }                  & y_{ua}y_{ub} \leq \epsilon,             \quad \forall u\in\Umc, a\neq b \in\Bmc \label{eq:num-ssa-gp-const1} \\
                                  & \sum_{u} y_{ub} \leq 1,                 \quad \forall b\in\Bmc                  \label{eq:num-ssa-gp-const2} \\
                                  & y_{ub} \in \Rbb_{++},                   \quad \forall u\in\Umc, b\in\Bmc.       \label{eq:num-ssa-gp-const3}
\end{align}
where \(\epsilon > 0\).
The constraint \eqref{num-ssa-gp-const1} may be re-tightened to that of the SSA problem by choosing \(\epsilon \to 0\).
The constraint \eqref{num-ssa-gp-const2} is relaxed from the equality constraint within \eqref{allocation-space}.
Finally, the optimization is performed over the positive orthant and specifically excludes variables \(y_{ub} = 0\).
This formulation is not directly convex, but may be easily convexified using a standard procedure detailed by Chiang~\cite{Chi2005}.

\begin{proposition}[\(\epsilon\)-SSA NUM is a GP]\label{prp:num-ssa-gp}
  The \(\epsilon\)-SSA NUM problem in \eqref{num-ssa-gp} is a geometric program (GP) when \(\alpha > 1\).
  When \(\alpha < 1\), the problem is not a GP.
\end{proposition}
\begin{IEEEproof}
  The objective function is a posynomial (sum of monomials), each with positive multiplicative constants when \(\alpha > 1\).
  The constraints \eqref{num-ssa-gp-const1} and \eqref{num-ssa-gp-const2} are posynomials, each with positive multiplicative constants and real exponents.
  When \(\alpha < 1\), the objective function is no longer a posynomial due to negative multiplicative constants.
\end{IEEEproof}

\begin{remark}[Loss of SSA in GP Formulation]
  The act of moving the summation outside of the utility function as in \eqref{objfun-manip} helped us create a GP that is convexifiable, but also prevents us from obtaining true SSA solutions when \(\alpha > 1\).
  So long as \(\sum_{b} r_{ub}y_{ub} \gg 0\), small individual allocations \(y_{ub} \sim 0\) have little effect on the original SSA NUM utility function (\lhs of \eqref{objfun-manip}).
  However, on the \rhs of \eqref{objfun-manip}, increasingly small individual allocations produce increasingly negative utilities.
  Due to this effect, solving the GP in \eqref{num-ssa-gp} tends to yield solutions that pull away from \(0\) on all dimensions.
  It then follows that the existence of a dominant allocation is prohibited by the very constraint meant to enforce it: \eqref{num-ssa-gp-const1}.
  As \(\epsilon \to 0\), individual allocations get even smaller due to \eqref{num-ssa-gp-const1} and we observe BS resource under-utilization (looseness of \eqref{num-ssa-gp-const2}).
\end{remark}

We include an example scenario demonstrating the combined effect of the objective function and constraints in \eqref{num-ssa-gp}.
\begin{example}[GP SSA with \(2\) BS and \(1\) MU]\label{ex:2bs1mu}
  Consider a small network with two BSs \(\Bmc = \{a,b\}\) and one MU \(\Umc = \{u\}\) and fairness parameter \(\alpha > 1\).
  Problem \eqref{num-ssa-gp} becomes:
  \begin{align}
    \min          \quad & \frac{1}{\alpha-1} \left((r_{ua}y_{ua})^{1-\alpha} + (r_{ub}y_{ub})^{1-\alpha}\right) \label{eq:gp-example} \\
    \textup{s.t.} \quad & y_{ua}y_{ub} -\epsilon \leq 0 \\
                        & y_{ua} - 1 \leq 0, \quad y_{ub} - 1 \leq 0 \\
                        & {-y_{ua}} \leq 0, \quad {-y_{ub}} \leq 0.
  \end{align}
  The Lagrangian of \eqref{gp-example} and its partial derivatives may be expressed:
  \begin{align}
    \Lmc(\ybf,\mubf) &= \frac{1}{\alpha-1} \left((r_{ua}y_{ua})^{1-\alpha} + (r_{ub}y_{ub})^{1-\alpha}\right) + \nonumber \\ 
    &\qquad \mu_1(y_{ua}y_{ub} - \epsilon) + \mu_{2a}(y_{ua}-1) + \mu_{2b}(y_{ub}-1) + \nonumber \\
    &\qquad \mu_{3a}(-y_{ua}) + \mu_{3b}(-y_{ub}) \\
    \frac{\partial \Lmc}{\partial y_{ua}} &= -r_{ua}^{1-\alpha}y_{ua}^{-\alpha} + \mu_1 y_{ub} + \mu_{2a} - \mu_{3a} \\
    \frac{\partial \Lmc}{\partial y_{ub}} &= -r_{ub}^{1-\alpha}y_{ub}^{-\alpha} + \mu_1 y_{ua} + \mu_{2b} - \mu_{3b},
  \end{align}
  with Lagrange multipliers \(\mubf = \{\mu_1, \mu_{2a}, \mu_{2b}, \mu_{3a}, \mu_{3b}\}\).
  From the Karush-Kuhn-Tucker (KKT) necessary conditions, we may characterize the optimal allocation by finding Lagrange multipliers \(\mubf^*\) that satisfy stationarity, primal feasibility, dual feasibility and complementary slackness are satisfied at any global minimum (\(\ybf^*\)).

  First, suppose that either \(y_{ua}^* \to 0\) or \(y_{ub}^* \to 0\).
  It follows that the objective function goes to \(\infty\), clearly an optimal solution requires \(y_{ua}, y_{ub} > 0\).

  Second, suppose that both \(y_{ua} = 1\) and \(y_{ub} = 1\).
  For small enough \(\epsilon\), we violate primal feasibility (\(y_{ua}y_{ub} \leq \epsilon\)).

  Third, suppose that both \(y_{ua}, y_{ub} \in (0,1)\).
  By complementary slackness, we have \(\mu_{2a} = \mu_{2b} = \mu_{3a} = \mu_{3b} = 0\).
  Solving the first order stationarity conditions for \(\mu_1\) yields:
  \begin{align}
    \mu_1 = \frac{r_{ua}^{1-\alpha}y_{ua}^{-\alpha}}{y_{ub}} &= \frac{r_{ub}^{1-\alpha}y_{ub}^{-\alpha}}{y_{ua}} \\
    r_{ua}^{1-\alpha}y_{ua}^{1-\alpha} &= r_{ub}^{1-\alpha}y_{ub}^{1-\alpha} \\
    r_{ua}y_{ua} &= r_{ub}y_{ub}.
  \end{align}
  Notice that \(\mu_1 > 0\).
  By complementary slackness, the primal constraint \(y_{ua}y_{ub} = \epsilon\) is tight.
  We may proceed to solve for the primal variables:
  \begin{align}
    y_{ua}^* &= \sqrt{\frac{\epsilon r_{ua}}{r_{ub}}} & y_{ub}^* &= \sqrt{\frac{\epsilon r_{ub}}{r_{ua}}}.
  \end{align}
  So long as \(\epsilon\) is chosen small enough, we have primal feasibility.

  Finally, \Wlog suppose that \(y_{ua} = 1\) and \(y_{ub} \in (0,1)\).
  The reduced problem can be expressed as:
  \begin{align}
    \min          \quad & \frac{1}{\alpha-1} (r_{ub}y_{ub})^{1-\alpha} \\
    \textup{s.t.} \quad & y_{ub} - \epsilon \leq 0 \\
                        & - y_{ub} \leq 0.
  \end{align}
  As \(\alpha>1\), the reduced problem is clearly solved by maximizing \(y_{ub} = \epsilon\), yielding solution \((y_{ua}^*, y_{ub}^*) = (1, \epsilon)\).

  Thus, there are three possible solutions to \eqref{gp-example}:
  \begin{equation}
    \ybf^* \in \left\{ \left( \sqrt{\frac{\epsilon r_{ua}}{r_{ub}}}, \sqrt{\frac{\epsilon r_{ub}}{r_{ua}}}\right), (1,\epsilon), (\epsilon,1)\right\}.
  \end{equation}

  We note that the objective function evaluated at all three possible optimal solutions grows as \(\epsilon \to 0\).
  However, the first solution is \(\Th{(1/\epsilon)^{(\alpha-1)/2}}\), while the latter two solutions are \(\Th{(1/\epsilon)^{\alpha-1}}\).
  It follows that the first solution yields the minimum for small \(\epsilon\).
  Thus, the GP solution under this network instance is highly under-allocated and the MU lacks a dominant allocation from a BS.
\end{example}

\begin{remark}[Re-capturing SSA]
  It seems that retaining the original SSA NUM objective function (\lhs of \eqref{objfun-manip}) would be sufficient to capture the SSA problem.
  However, the original objective function, while a composition of posynomials \(\{\sum_{b} r_{ub}y_{ub}, \forall u\in\Umc\}\) with a posynomial, is not a generalized GP due to negative exponents in the outer posynomial \cite{Chi2005}.
  Although we can not prove that such a problem is not convexifiable, we do not currently know how to convexify it.
\end{remark}

%% file: num-ssa-unialloc.tex
In \secref{num-ssa-optalloc}, we permitted general BS allocation schemes \(\ybf \in \Ymc\).
It is conceivable that the additional overhead incurred by BSs to monitor instantaneous downlink rates and adjust resources may not be desirable, so we propose to study the SSA NUM problem restricted to a uniform allocation scheme, wherein each BS shares its resource uniformly among all MUs with which it is associated.
Under this allocation scheme, the resulting NUM is also completely combinatorial in nature and is given by \prpref{num-ssa-unialloc}.
\begin{proposition}[SSA NUM Problem w/ Uniform Allocation]\label{prp:num-ssa-unialloc}
  Under a uniform allocation scheme, the SSA NUM problem \eqref{num-ssa}, denoted \(\fSU\), is given by the third column of \tabref{num-ssa}.
\end{proposition}
\begin{IEEEproof}
  See proof in \prfref{num-ssa-unialloc}.
\end{IEEEproof}

Observe the optimal and uniform allocation problems are identical at $\alpha=1$.

Given the difficulty of the combinatorial problem, we again consider an integer relaxation of the SSA NUM problem under uniform allocation schemes, which may be derived from \tabref{num-ssa} by replacing \(\zbf \in \Zmc\) with \(\xbf \in \Xmc\), the feasible RSSA association space.
Due to the fact that \(\fRSU\) is an integer relaxation of \(\fSU\), we immediately know that \(\fSU \leq \fRSU\), but \thmref{num-ssa-unialloc-int-gap} shows that there is no integrality gap in the SSA NUM problem for \(\alpha\in[0,2]\).
In this regime of \(\alpha\), \(\fRSU\) may be attained via integer solutions that simultaneously attain \(\fSU\).
We summarize our findings on the integrality gap in \tabref{num-ssa-int-convex}.
\begin{theorem}[Integrality Gap of \(\fSU\)]\label{thm:num-ssa-unialloc-int-gap}
  When \(\alpha\in[0,2]\), the integrality gap associated with the SSA NUM problem \(\fSU\) is equal to \(1\):
  \begin{equation}
    \fSU(\zbf^*) / \fRSU(\xbf^*) = 1, \quad \alpha \in [0,2].
  \end{equation}
\end{theorem}
\begin{IEEEproof}
  See proof in \prfref{num-ssa-unialloc-int-gap}.
\end{IEEEproof}

While integer relaxation is a valid approach to solving the SSA NUM problem, we again find that the relaxed problem \(\fRSU\) generally results in non-convex problems and organize these results in \tabref{num-ssa-int-convex}.
\begin{proposition}[Non-convexity of \(\fRSU\)]\label{prp:num-rssa-unialloc-convexity}
  \(\fRSU\) is non-convex for \(\alpha \geq 0, \alpha \neq 1\) and is convex for \(\alpha=1\).
\end{proposition}
\begin{IEEEproof}
  The proof mirrors that of \prpref{num-rssa-optalloc-convexity}.
\end{IEEEproof}

In summary, the SSA NUM problems exhibit unit integrality gap for $\alpha \in [0,2]$ and non-convexity for $\alpha \neq 1$, for both optimal and uniform resource allocations.

%% file: num-ssa-algos.tex
In this section, we present four greedy algorithms to obtain feasible solutions to the SSA NUM \eqref{num-ssa}.
\algref{cga} provides a centralized approach, while \algref{lga} and \algref{lgan} both make use of a localized approach, with the corresponding algorithms running on each MU  and on each BS.
In \algref{lga}, MUs and BSs exchange instantaneous rate information which may be used in association and allocation decisions, while \algref{lgan} captures the scenario in which instantaneous rate information is not shared throughout the network.
The fourth algorithm applies a greedy rounding of MSA allocations to obtain feasible SSA associations.
For all four algorithms, either the optimal or uniform allocation policies may be paired with the feasible SSA associations.
Finally, it will be useful to define \(\oneub\) as adding an association of MU $u$ with BS $b$, and to denote by \(\Pmc\)  the set of (currently) unassociated MUs.

\subsection{Centralized Greedy Algorithm (CGA)}\label{sec:cga}

\algref{cga} chooses an additional association in each while loop iteration that effectively results in the largest increase in network utility:
\begin{equation}\label{eq:cga-argmax}
  \argmax_{\substack{v\in\Pmc,\\ a\in\Bmc}} \fSSA(\zbfva) = \argmax_{\substack{v\in\Pmc,\\ a\in\Bmc}} \fSSA(\zbfva) - \fSSA(\zbf).
\end{equation}
Note, the \(\argmax\) expression above may be simplified further by considering that the change in network utility due to MU \(v\) is localized to the change in utility at BS \(a\):
\begin{align}
  \fSSA(\zbfva) - &\fSSA(\zbf) \\
  &\stackrel{(a)}{=} \sum_{ub} \utility{r_{ub}y_{ub}^*(\zbfva)}(\zbfva)_{ub} \nonumber \\
  &\qquad- \sum_{ub} \utility{r_{ub}y_{ub}^*(\zbf)}z_{ub} \\
  &\stackrel{(b)}{=} \sum_{u} \utility{r_{va}y_{va}^*(\zbfva)}(\zbfva)_{ua} \nonumber \\
  &\qquad- \sum_{u} \utility{r_{ua}y_{ua}^*(\zbf)}z_{ua}, \label{eq:cga-utildiff}
\end{align}
where
\emph{(a)} expands \(\fSSA\) using \eqref{objfun-manip}, and
\emph{(b)} cancels identical terms corresponding to all BSs other than \(a\).
Finally, ties in the \(\argmax\) are broken arbitrarily to determine the new association.

\begin{algorithm}[t!]
  \caption{Centralized Greedy Algorithm (CGA)}
  \label{alg:cga}
  \begin{algorithmic}[1]
    \Function{CGA}{}
      \State \(\zbf \gets \mathbf{0}; \quad \Pmc \gets \Umc\)
      \While{\(\Pmc \neq \emptyset\)}
        \State \((u,b) \gets \argmax_{v\in\Pmc, a\in\Bmc} \fSSA(\zbfva)\)
        \State \(z_{ub} \gets 1; \quad \Pmc \gets \Pmc \setminus \{u\}\)
      \EndWhile
      \State \Return \(\zbf\)
    \EndFunction
  \end{algorithmic}
\end{algorithm}

\begin{algorithm}[t!]
  \caption{Localized Greedy Algorithm (LGA)}
  \label{alg:lga}
  \begin{algorithmic}[1]
    \Function{LGA}{}
      \State \(\zbf \gets \mathbf{0}; \quad \Pmc \gets \Umc; \quad R_b \gets \emptyset, \forall b\in\Bmc\)
      \While{\(\Pmc \neq \emptyset\)}
        \For{\(\forall u\in\Pmc\)}
          \State \(b \gets \argmax_{a\in\Bmc} \fSSA(\zbfua)\)
          \State \(R_b \gets R_b \cup \{u\}\)
        \EndFor
        \For{\(\forall b\in\Bmc\)}
          \If{\(R_b \neq \emptyset\)}
            \State \(u \gets \argmax_{v \in R_b} \fSSA(\zbfvb)\)
            \State \(z_{ub} \gets 1; \quad \Pmc \gets \Pmc \setminus \{u\}; \quad R_b \gets \emptyset\)
          \EndIf
        \EndFor
      \EndWhile
      \State \Return \(\zbf\)
    \EndFunction
  \end{algorithmic}
\end{algorithm}

\begin{algorithm}[t!]
  \caption{LGA with No Rate Information (LGAN)}
  \label{alg:lgan}
  \begin{algorithmic}[1]
    \Function{LGAN}{}
      \State \(\zbf \gets \mathbf{0}; \quad \Pmc \gets \Umc; \quad R_b \gets \emptyset, \forall b\in\Bmc\)
      \State \(\kappa_b \gets 0, \forall b\in\Bmc\)
      \While{\(\Pmc \neq \emptyset\)}
        \For{\(\forall u\in\Pmc\)}
          \State \(b \gets \argmax_{a\in\Bmc} \utility{\frac{r_{ua}}{\kappa_a+1}}\)
          \State \(R_b \gets R_b \cup \{u\}\)
        \EndFor
        \For{\(\forall b\in\Bmc\)}
          \If{\(R_b \neq \emptyset\)}
            \State \(u \gets \argmax_{v \in R_b} 1\)
            \State \(z_{ub} \gets 1; \quad \Pmc \gets \Pmc \setminus \{u\}; \quad R_b \gets \emptyset\)
            \State \(\kappa_b \gets \kappa_b + 1\)
          \EndIf
        \EndFor
      \EndWhile
      \State \Return \(\zbf\)
    \EndFunction
  \end{algorithmic}
\end{algorithm}

\subsection{Localized Greedy Algorithm (LGA)}\label{sec:lga}

Noting that the change in network utility due to a single additional association may be localized, we are motivated to construct a similar algorithm \algref{lga} with a split-phase while loop.
In the first phase (lines 4-6), each unassociated MU \(u\) independently requests to associate with a BS that effectively maximizes the increase in localized utility.
Requests for each BS \(b\) are stored in \(R_b\).
In the second phase (lines 7-10), each BS \(b\) independently grants association to the requesting MU that effectively maximizes the increase in localized utility at \(b\).
Upon granting an association request, the set of unassociated MUs is reduced, and BS \(b\) resets \(R_b\).

In order for MUs to compute the \(\argmax\) in line 5 of \algref{lga}, each MU must know the instantaneous rates of assigned MUs, which may be accomplished by requiring \emph{i)} each MU to transmit its instantaneous rate along with its association request, and \emph{ii)} each BS to advertise the instantaneous rate of the newly associated MU after each association round.
Similarly, the \(\argmax\) in line 9 of \algref{lga} requires each BS to track the instantaneous rates of associated MUs which may be included with each MU's association request.

\subsection{LGA with No Rate Information (LGAN)}\label{sec:lgan}

Finally, we are motivated to design a similar localized algorithm, \algref{lgan}, that does not require the sharing of instantaneous rate information between MUs and BSs.
In the absence of rate information, we assume each BS allocates its resources uniformly across its associated MUs; this allocation means the effective state of each BS is the number of associated MUs, denoted \(\kappa_b\), and the algorithm necessitates that each MU track these congestion counts at each BS.
In the first phase (lines 5-7), each unassociated MU \(u\) independently requests to associate with a BS that maximizes the MU's individual utility.
Requests for each BS \(b\) are stored in \(R_b\).
In the second phase (lines 8-12), each BS \(b\) independently grants association to an arbitrarily chosen requesting MU.
Upon granting an association request, the set of unassociated MUs is reduced, BS \(b\) resets \(R_b\) and increases its congestion count.

In order for MUs to compute the \(\argmax\) in line 6 of \algref{lga}, each MU must know the congestion count of each BS, which may be accomplished by requiring \emph{i)} each BS to advertise granted associations, and \emph{ii)} each MU to increment locally stored congestion counts.
The \(\argmax\) in line 10 of \algref{lga} requires no rate information.

\subsection{MSA NUM Rounding (MSARnd)}\label{sec:msa-rounding}

Given an optimal MSA allocation \(\ybf^*\) to the MSA NUM problem, \(\fMSA\), we wish to convert it to a feasible SSA association \(\zbf\) and allocation \(\ybf\).
We choose to associate each MU \(u\) to the BS \(b(u)\) that offers the largest rate (breaking ties arbitrarily):
\begin{align}
  z_{ub}(\ybf^*) &= \mathbf{1}\{b = \hat{b}(u)\}, & \hat{b}(u) &\in \argmax_{b\in\Bmc} r_{ub}y^*_{ub}.
\end{align}
Having obtained a feasible association \(\zbf\) in this manner from $\ybf^*$, we may then compute the optimal allocation \(\ybf(\zbf)\) associated with $\zbf$ using \prpref{num-ssa-solution}; observe $\ybf(\zbf(\ybf^*)) \neq \ybf^*$.
As with the other algorithms presented in this section, we may bound the optimal value \((\hat{\ybf}^*, \hat{\zbf}^*)\) of \(\fSSA\) using both a feasible SSA association \(\zbf\) and the optimal MSA allocation \(\ybf^*\) for the lower and upper bounds respectively:
\begin{equation}
  \fSSA(\ybf(\zbf),\zbf) \leq \fSSA(\hat{\ybf}^*,\hat{\zbf}^*) \leq \fMSA(\ybf^*).
\end{equation}

%% file: results.tex
In this section, we study the performance of the four SSA algorithms proposed in \secref{num-ssa-algos}, using the MSA NUM problem as a baseline network \(\alpha\)-utility measure.
The achievable utility under the MSA NUM problem serves as an upper bound on the achievable utility under the SSA NUM problem and any of its associated algorithms in \secref{num-ssa-algos}.

\begin{figure*}[t!]
  \centering%
  \includegraphics[width=0.5\textwidth]{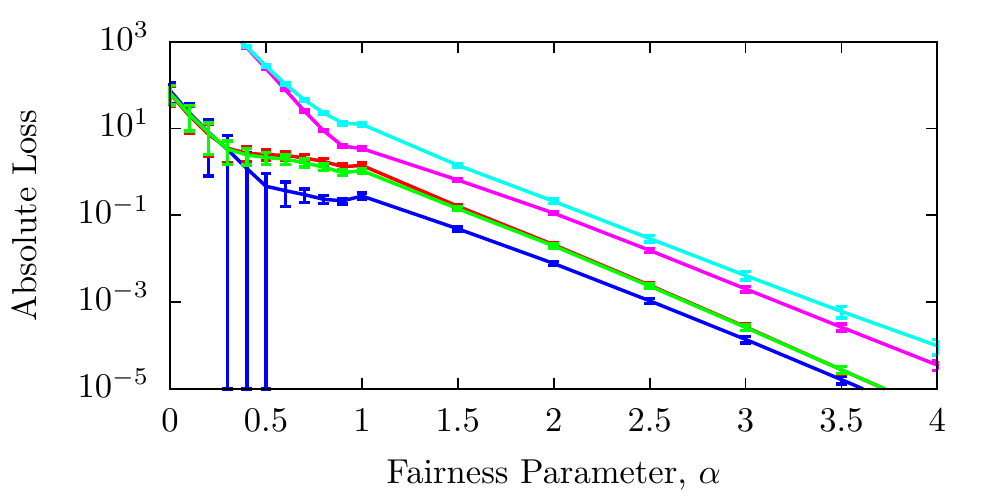}%
  \includegraphics[width=0.5\textwidth]{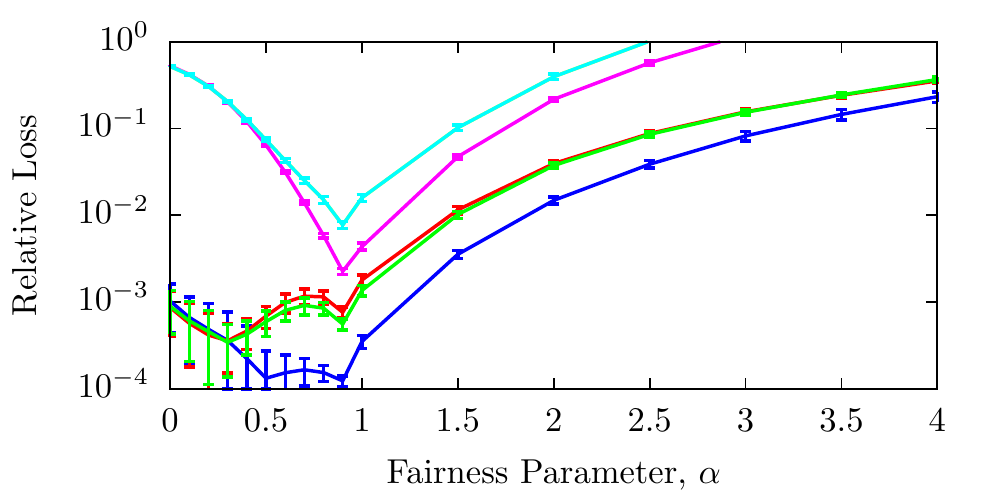}\\%
  \includegraphics[width=0.5\textwidth]{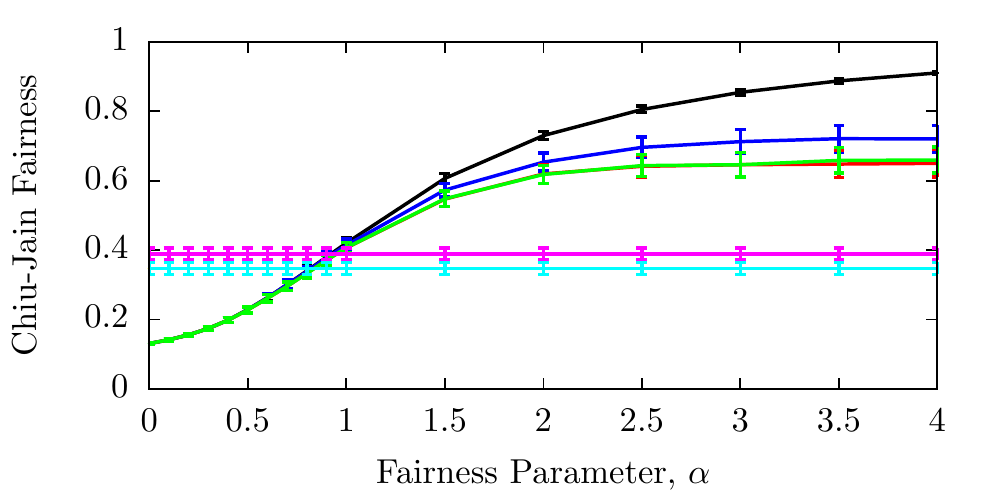}%
  \includegraphics[width=0.5\textwidth]{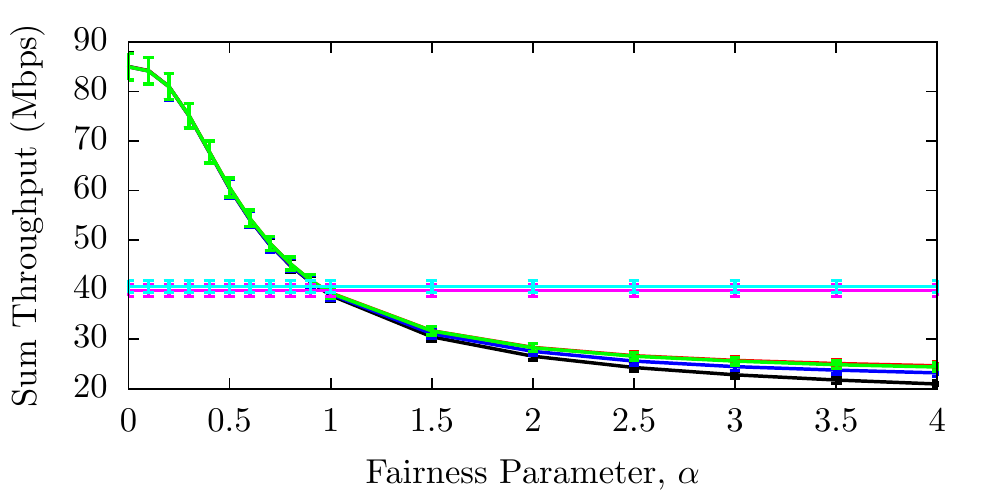}\\%
  \includegraphics[width=0.75\textwidth]{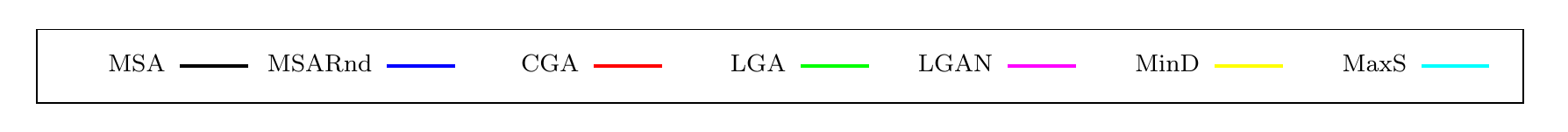}%
  \caption{%
    The absolute (top-left) and relative (top-right) loss in utility incurred by a collection of heuristics for the SSA NUM problem over the optimal MSA NUM problem.
    The Chiu-Jain fairness (bottom-left) and network sum throughput (bottom-right) associated with the MSA NUM problem and a collection of heuristics for the SSA NUM problem.
    Results are generated from \(100\) independent samples with \(95\%\) confidence intervals.
    Each network consists of \(100\) MUs and \(20\) BSs with positions generated uniformly at random within a square arena with a side length of \(1000\) meters.
    Each BS transmits at \(1000\) mW with a signal bandwidth of \(1.2\) MHz, background noise is assumed to be \(-90\) dBm, and the pathloss constant is set to \(3\).
  }%
  \label{fig:secV}%
\end{figure*}

As shown in \figref{secV}, the absolute (top-left) and relative (top-right) loss in sum-user network utility for the SSA solutions obtained by the four algorithms (\MSARnd, \CGA, \LGA, and \LGAN) relative to the MSA solution is shown as a function of the $\alpha$ parameter.
Additionally, we include a heuristic where each MU associates with the min distance BS (denoted \MinD) and a heuristic where each MU associates with its max SINR BS (denoted \MaxS).
Given the association $\zbf$, the three algorithms \MSARnd, \CGA, and \LGA are configured to generate optimal allocations \(\ybf^*\), as studied in \secref{num-ssa-optalloc}, while the three heuristics \LGAN, \MinD, and \MaxS employ uniform allocations \(\ybf\), as studied in \secref{num-ssa-unialloc}.  The justification behind this decision is that \MSARnd, \CGA, and \LGA employ instantaneous rate information in forming the association $\zbf$, so they may as well use this same information in forming the allocation $\ybf^*$.  Conversely, \LGAN, \MinD, and \MaxS do not employ rate information in forming the association $\zbf$, and so it is reasonable for the BS to use a uniform allocation $\ybf$.

We observe that the algorithms making use of optimal allocations (\MSARnd, \CGA, \LGA) incur lower relative and absolute losses in \(\alpha\)-utility.
\MSARnd, which uses the optimal MSA allocations as a starting point, outperforms all other SSA algorithms, but incurs the cost of requiring a convex problem solver.
Both simple heuristics (\MinD and \MaxS) perform the worst, as association decisions are made locally without any congestion information.

When attempting to draw conclusions about algorithm performance as a function of the \(\alpha\)-utility parameter, we observe some care is needed.
Absolute and relative losses tend to be large for small and large \(\alpha\), respectively.
The unitless nature of \(\alpha\)-utility also presents difficulty when comparing utilities under two different values of \(\alpha\).
Thus, we again invoke the Chiu-Jain fairness measure \cite{JaiChiHaw1984}.
However, in this application, we measure the equity of the sum rates across users:
\begin{equation}
  \Jmc([R_u(\ybf,\zbf), \forall u\in\Umc]) = \frac{\left(\sum_{u} R_u(\ybf,\zbf)\right)^2}{|\Umc| \sum_{u} \left(R_u(\ybf,\zbf)\right)^2},
\end{equation}
The Chiu-Jain fairness measure used in this way ranges over the interval \([1/\Umc,1]\); the endpoints of this interval capture the scenarios $i)$ where only one MU receives a non-zero sum rate (the least equitable, \(\Jmc_u = 1/\Bmc\)), and $ii)$ where all MUs receive an equal sum rate (the most equitable, \(\Jmc_u = 1\)).
We expect fairness to be acheived at the cost of the network's sum throughput, so \figref{secV} plots both fairness (bottom-left) and throughput (bottom-right) as a function of the \(\alpha\)-utility parameter.
We observe that \(\alpha\) does indeed act as a proxy for controlling fairness in the resulting optimal MSA allocations; when \(\alpha < 1\) throughput maximization is emphasized over fairness, while \(\alpha > 1\) results in more fair allocations achieving lower throughput.
Due to the restriction in feasible allocations associated with the SSA NUM problem, we observe that the SSA NUM algorithms do not reproduce the same fairness as do optimal MSA allocations; in fact, the SSA NUM algorithm fairness curves (\MSARnd, \CGA, and \LGA) appear to be limited as \(\alpha\) grows large.
Again, \MSARnd appears to be the most equipped to replicate the fairness of optimal MSA NUM allocations, while \LGAN, \MinD, and \MaxS are all insensitive to \(\alpha\).

\begin{figure*}[t!]
  \centering%
  \includegraphics[width=0.5\textwidth]{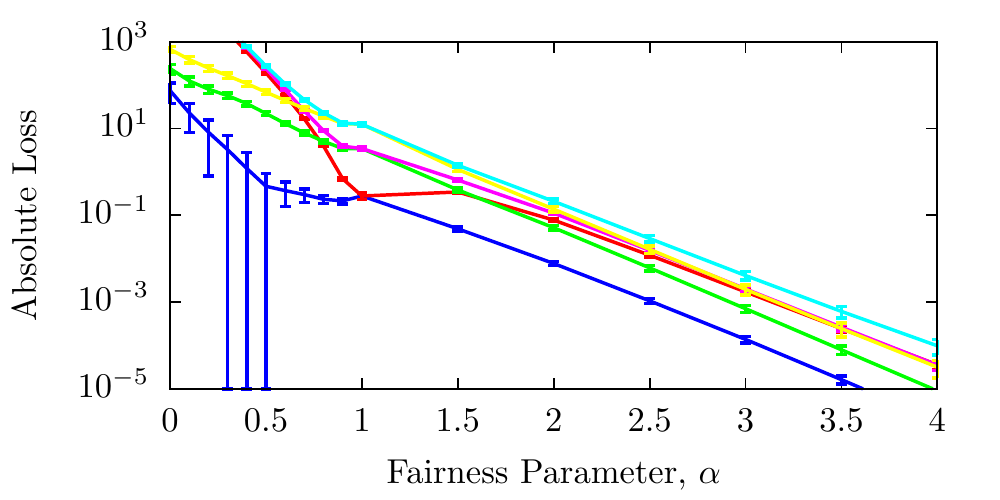}%
  \includegraphics[width=0.5\textwidth]{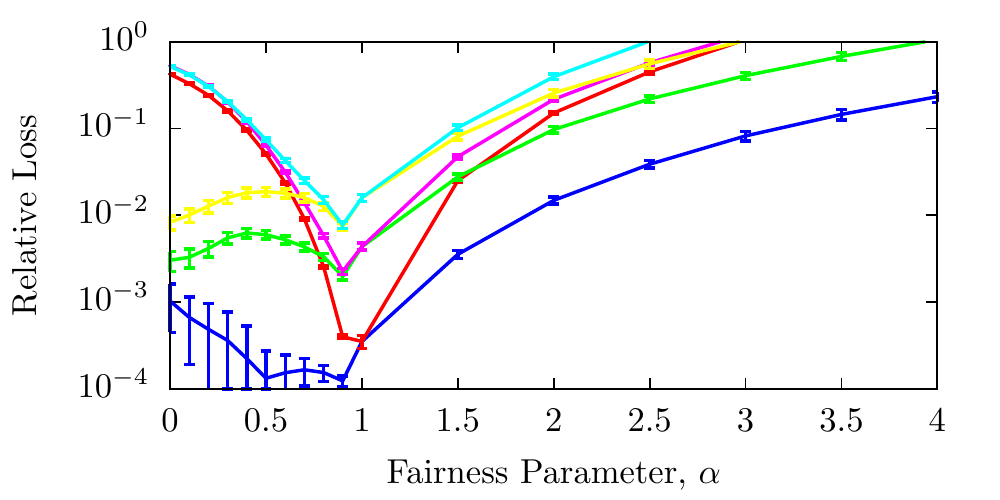}\\%
  \includegraphics[width=0.5\textwidth]{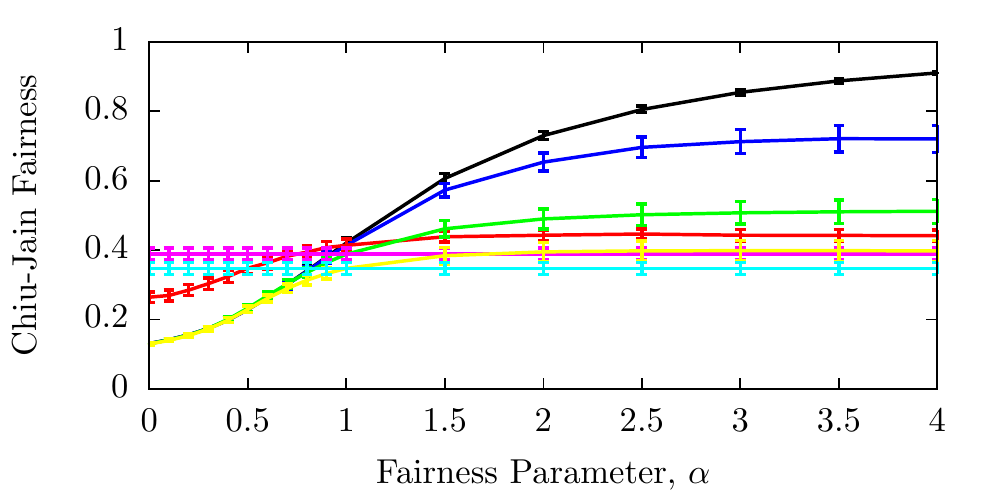}%
  \includegraphics[width=0.5\textwidth]{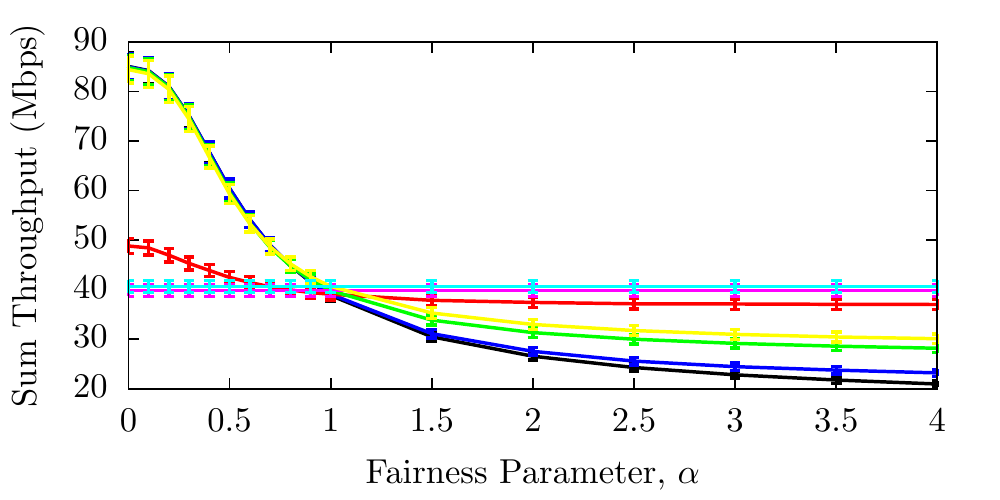}\\%
  \includegraphics[width=0.85\textwidth]{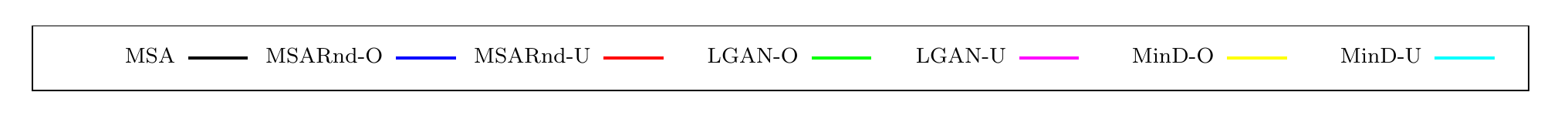}%
  \caption{%
    The absolute (top-left) and relative (top-right) loss in utility incurred by a collection of heuristics for the SSA NUM problem over the optimal MSA NUM problem.
    The Chiu-Jain fairness (bottom-left) and network sum throughput (bottom-right) associated with the MSA NUM problem and a collection of heuristics for the SSA NUM problem.
    Each of the SSA heuristics generate a feasible association \(\zbf\), and are combined with either optimal or uniform BS resource allocations \(\ybf(\zbf)\), denoted O and U, resp.
    With the exception of a fixed MU:BS ratio of \(5:1\), the results are generated with the same parameters as \figref{secV}.
  }%
  \label{fig:secV2}%
\end{figure*}

In \figref{secV2}, we explore the effect of BS resource allocation schemes with a subset of the SSA algorithms used in \figref{secV}.
As expected, for a given SSA algorithm, optimally allocating BS resources yields lower utility losses (top-left and top-right) than uniformly allocating BS resources.
However, when \(\alpha = 1\), the uniform allocation scheme is optimal (\remref{num-ssa-solution}) and we observe no additional utility loss for each of the three SSA algorithms shown (\MSARnd, \LGAN, and \MinD).
With respect to the fairness (bottom-left) of the SSA heuristics, we note that optimizing BS resource allocations can have a dramatic impact on the achievable fairness range (\cf \MSARnd-O and \MSARnd-U).
Even for simpler SSA heuristics like \MinD, we note an improvement in Chiu-Jain fairness as a result of solving for the optimal over uniform allocations.
Finally, while both \MinD and \MaxS under uniform allocations are insensitive to \(\alpha\), \MSARnd under uniform allocations still responds to \(\alpha\) via the rounding process to obtain a feasible association \(\zbf\).

%% file: appendix.tex
\subsection{\prpref{num-ssa-solution} (Solution of \texorpdfstring{\(\fSSA(\zbf,b)\)}{fSSA(z,b)})}\label{prf:num-ssa-solution}

\begin{IEEEproof}
  When \(\alpha = 0\), the objective function becomes:
  \begin{equation}
    \max \sum_{u\in\Umc_{b}} r_{ub}y_{ub},
  \end{equation}
  which is clearly maximized by allocating all BS resources to MUs with the largest instantaneous downlink rate.

  When \(\alpha=1\), we call upon the results of Ye \ea~\cite[Prp.~1]{YeRonChe2013}, who showed that the optimal allocation is uniform.

  We now focus on \(\alpha > 0, \alpha \neq 1\).
  The Lagrangian of \(\fSSA(\zbf,b)\) and its partial derivatives may be expressed:
  \begin{align}
    \Lmc(\ybf,\lambda_b,\mubf)            &= - \sum_{u\in\Umc_{b}} U_{\alpha}(r_{ub}y_{ub}) + \lambda_b\left(\sum_{u\in\Umc_b} y_{ub} - 1\right) \nonumber \\
    & \qquad + \sum_{u\in\Umc_b} \mu_{u} (- y_{ub}) \\
    \frac{\partial \Lmc}{\partial y_{ub}} &= -r_{ub}^{1-\alpha}y_{ub}^{-\alpha} + \lambda_b - \mu_u, \quad \forall u\in\Umc_b,
  \end{align}
  with Lagrange multipliers \(\lambda_b \in \Rbb\) and \(\mubf = \{\mu_u, u \in \Umc_b\}\).
  By convexity of \(\fSSA(\zbf,b)\), Karush-Kuhn-Tucker (KKT) conditions are necessary and sufficient for optimality.
  By KKT, there must exist Lagrange multipliers \(\lambda_b^*\) and \(\mubf^*\) such that
  stationarity \eqref{num-ssa-stationarity},
  primal feasibility \eqref{num-ssa-primal-feas},
  dual feasibility \eqref{num-ssa-dual-feas},
  and complementary slackness \eqref{num-ssa-comp-slack} are satisfied at any global minimum (\(\ybf^*\)):
  \begin{align}
    &\frac{\partial \Lmc}{\partial y_{ub}} = -r_{ub}^{1-\alpha}y_{ub}^{-\alpha} + \lambda_b - \mu_u = 0, \quad \forall u\in\Umc_b \label{eq:num-ssa-stationarity} \\
    &\begin{cases} \sum_{u\in\Umc_b} y_{ub} - 1 = 0 
      \\ -y_{ub} \leq 0, \quad \forall u\in\Umc_b
    \end{cases} \label{eq:num-ssa-primal-feas} \\
    &\mu_u \geq 0, \quad \forall u\in\Umc_b \label{eq:num-ssa-dual-feas} \\
    &\mu_u(-y_{ub}) = 0, \quad \forall u\in\Umc_b, \label{eq:num-ssa-comp-slack}
  \end{align}

  We now characterize the optimal solution using the KKT conditions.
  Assume that all \(y_{ub}^* > 0, \forall u\in\Umc_b\), which partially satisfies primal feasibility \eqref{num-ssa-primal-feas}.
  Additionally setting \(\mu_{u}^* = 0\) satisfies dual feasibility \eqref{num-ssa-dual-feas} and complementary slackness \eqref{num-ssa-comp-slack}.
  Next, the resulting stationarity conditions yield:
  \begin{equation}\label{eq:ssa2-proof1}
    y_{ub}^* = \frac{r_{ub}^{(1-\alpha)/\alpha}}{(\lambda_b^*)^{1/\alpha}}.
  \end{equation}

  Substituting the above into the remaining primal feasibility conditions in \eqref{num-ssa-primal-feas}, we obtain:
  \begin{equation}
    \lambda_b^* = \left( \sum_{u\in\Umc_b} r_{ub}^{(1-\alpha)/\alpha} \right)^{\alpha}.
  \end{equation}

  Substituting \(\lambda_b^*\) back into \eqref{ssa2-proof1}, we obtain the desired form of \(y_{ub}^*\):
  \begin{equation}
    y_{ub}^* = \frac{s_{ub}}{\sum_{v\in\Umc_b} s_{vb}},
  \end{equation}
  using \(s_{ub} = r_{ub}^{(1-\alpha)/\alpha}\) as convenience variables.
  The remaining primal feasibility conditions \eqref{num-ssa-primal-feas} are easily verified.

  Finally, when \(\alpha \to \infty\), we note that \(s_{ub} \to r_{ub}^{-1}\).
\end{IEEEproof}

\subsection{\prpref{num-ssa-sensitivity} (Sensitivity of \(y_{ub}^*\) to Instantaneous Rates)}\label{prf:num-ssa-sensitivity}

\begin{IEEEproof}
  \begin{align}
    \frac{\partial y_{ub}^*}{\partial r_{ub}}
    &= \frac{\partial}{\partial r_{ub}} \frac{s_{ub}}{\sum_{v\in\Umc_b} s_{vb}}
    = \frac{\frac{\partial s_{ub}}{\partial r_{ub}} \left(\sum_{v\in\Umc_b\setminus u} s_{vb} \right)  }{\left(\sum_{v\in\Umc_b} s_{vb}\right)^2}
  \end{align}

  \begin{align}
    \frac{\partial s_{ub}}{\partial r_{ub}}
    &= \frac{\partial }{\partial r_{ub}} r_{ub}^{(1-\alpha)/\alpha}
    =  \frac{1-\alpha}{\alpha} r_{ub}^{1/\alpha-2}
  \end{align}

  \begin{align}
    \frac{\partial y_{ub}^*}{\partial r_{ub}}
    = \frac{\frac{1-\alpha}{\alpha} r_{ub}^{1/\alpha-2} \left(\sum_{v\in\Umc_b\setminus u} s_{vb} \right)  }{\left(r_{ub}^{1/\alpha-1} +  \sum_{v\in\Umc_b\setminus u} s_{vb}\right)^2}
  \end{align}

  The partial derivative clearly changes sign at \(\alpha=1\).
  When \(\alpha=1\), we observe that the optimal allocation policy is independent of instantaneous rates.
\end{IEEEproof}

\subsection{\prpref{num-ssa-optalloc} (SSA NUM Problem w/ Optimal Allocation)}\label{prf:num-ssa-optalloc}

\begin{IEEEproof}
  When \(\alpha \rightarrow \infty\), \eqref{num-ssa} becomes a minimax problem:
  \begin{align}
    &\argmax_{\zbf} \lim_{\alpha\rightarrow\infty} \sum_{u} \utility{R_{u}(\ybf(\zbf))} \\
    &\quad = \argmin_{\zbf} \lim_{\alpha\rightarrow\infty} \frac{1}{\alpha-1} \sum_{u} \left(\sum_{b} r_{ub}y_{ub}(\zbf)\right)^{\alpha-1} \\
    &\quad \stackrel{(a)}{=} \argmin_{\zbf} \max_{u} \sum_{b} z_{ub} \left(r_{ub}^{-1} + \sum_{v}^{v\neq u} r_{vb}^{-1} z_{vb} \right)^{-1} \\
    &\quad = \argmax_{\zbf} \min_{u} \left( \sum_{b} z_{ub} \left(r_{ub}^{-1} + \sum_{v}^{v\neq u} r_{vb}^{-1} z_{vb} \right)^{-1} \right)^{-1},
  \end{align}
  where in (a) we apply the approximation for \(\ybf(\zbf)\) as \(\alpha \to \infty\) given in \prpref{num-ssa-solution} before taking the limit.

  For finite \(\alpha > 0\), we have:
  \begin{align}
    \sum_{u} \utility{ R_{u}(\ybf(\zbf)) }
    &= \sum_{u} \utility{ \sum_{b} r_{ub}y_{ub}(\zbf) } \\
    &= \sum_{ub} z_{ub} \utility{ r_{ub}y_{ub}(\zbf) },
  \end{align}
  where we move the summation over \(b\) outside of the utility function as \(y_{ub}(\zbf)\) is active for exactly one summand.

  Finally, when \(\alpha = 0\), we continue the above:
  \begin{align}
    \sum_{u} \utility{ R_{u}(\ybf(\zbf)) }
    &= \sum_{b} \sum_{u} r_{ub}y_{ub}(\zbf)z_{ub} \\
    &= \sum_{b} \max_{u} r_{ub}z_{ub},
  \end{align}
  noting that the optimal solution \(\ybf(\zbf)\) from \prpref{num-ssa-solution} has each BS \(b\) allocate all resources to associated users with the highest rate.
\end{IEEEproof}

\subsection{\thmref{num-ssa-optalloc-int-gap} (Integrality Gap of \texorpdfstring{\(\fSO\)}{fSO})}\label{prf:num-ssa-optalloc-int-gap}

\begin{IEEEproof}
  Let \(\xbf^*\) be an optimal solution to \(\fRSO\).
  Consider the following randomized rounding scheme from \(\Xmc\) to \(\Zmc\).
  Let \(\Xbf = (X_u, u\in\Umc)\) represent a random feasible SSA solution where each \(X_u \in \{1,\dots,|\Bmc|\}\) indicates the index of the BS to which MU \(u\) is assigned.
  Let each \(X_u\) be independently chosen via a distribution induced by the optimal solution \(\xbf^*\): \(\Prob{X_u = b} = x_{ub}^*\).
  Next, let \(\Zbf = (Z_{ub}, u\in\Umc,b\in\Bmc)\) be indicator \rv's formed from \(\Xbf\): \(Z_{ub} = \mathbf{1}_{\{X_u = b\}}\), and let \(Y_{ub} = \sum_{v}^{v\neq u} Z_{vb}\).
  We now bound the expected value of \(\fRSO(\Zbf)\).

  For the case of \(\alpha=0\), we carry out the steps explicitly:
  \begin{align}
    &\Expect{f_{\alpha=0}^{\textup{RSO}}(\Zbf)} = \Expect{\sum_{b} \max_{u} r_{ub}Z_{ub}} \\
    &\qquad \stackrel{(a)}{\geq} \sum_{b} \max_{u} r_{ub}\Expect{Z_{ub}} \\
    &\qquad \stackrel{(b)}{=} \sum_{b} \max_{u} r_{ub}x_{ub}^* \\
    &\qquad = f_{\alpha=0}^{\textup{RSO}}(\xbf^*),
  \end{align}
  where we have used
  \emph{(a)} the independence of \(Z_{ub}\),
  \emph{(b)} Jensen's inequality and the convexity of \(\max\).

  Similarly, for the case of \(\alpha=1\), we carry out the steps explicitly:
  \begin{align}
    &\Expect{f_{\alpha=1}^{\textup{RSO}}(\Zbf)} = \Expect{\sum_{ub} Z_{ub} \left(\log\!\left( r_{ub} \right) - \log\!\left(1 \!+\! Y_{ub}\right) \right)} \\
    &\qquad \stackrel{(a)}{=} \sum_{ub} \Expect{Z_{ub}} \left(\log\!\left( r_{ub} \right) - \Expect{\log\!\left(1 + Y_{ub}\right)} \right) \\
    &\qquad \stackrel{(b)}{\geq} \sum_{ub} \Expect{Z_{ub}} \left(\log\!\left( r_{ub} \right) - \log\!\left(1 + \Expect{Y_{ub}}\right) \right) \\
    &\qquad \stackrel{(c)}{=} \sum_{ub} x_{ub}^* \left(\log\!\left( r_{ub} \right) - \log\!\left(1 + \sum_{v}^{v\neq u} x_{vb}^*\right) \right) \\
    &\qquad = f_{\alpha=1}^{\textup{RSO}}(\xbf^*),
  \end{align}
  where we have used
  \emph{(a)} the independence of \(Z_{ub}\) and \(Y_{ub}\),
  \emph{(b)} Jensen's inequality and the concavity of \(\log(1+Y_{ub})\), and
  \emph{(c)} \(\Expect{Y_{ub}} = \sum_{v}^{v\neq u} \Expect{X_{vb}}\).

  To generalize this argument over \(\alpha\), we define \(g_{\alpha}(y)\):
  \begin{equation}
    g_{\alpha}(y) \equiv \begin{cases}
      \frac{1}{1-\alpha} y^{\alpha-1},    & \alpha \neq 1 \\
      \log(1/y),                          & \alpha = 1
    \end{cases},
  \end{equation}
  and note that over \(y \in \Rbb_+\), it is convex for \(\alpha \in (0,2)\), linear for \(\alpha=2\), and concave for \(\alpha > 2\).
  %
  %
    We also generalize \(Y_{ub} = \sum_{v}^{v\neq u} d_{uvb}(\alpha) Z_{vb}\) where:
    \begin{equation}
      d_{uvb}(\alpha) = \begin{cases}
        \left(\frac{r_{vb}}{r_{ub}}\right)^{(1-\alpha)/\alpha} & \alpha \neq 1 \\
        1, & \alpha = 1
      \end{cases}.
    \end{equation}
    Note that the composition of \(g(y)\) with the affine mapping \(y = 1 + Y_{ub} = 1+\sum_{v}^{v\neq u} d_{uvb}(\alpha) Z_{vb}\) preserves the concavity/convexity of \(g\) \cite[\S{}3.2.2]{BoyVan2004}.
  Using Jensen's inequality on \(g_{\alpha}(y)\) and expanding \(y\) we obtain the following inequalities:
  \begin{align}
    \Expect{\!\frac{(1\!+\!Y_{ub})^{\alpha\!-\!1}}{1\!-\!\alpha}\!}  &\geq \frac{(1\!+\!\Expect{Y_{ub}})^{\alpha\!-\!1}}{1\!-\!\alpha},   & \alpha& \in (0,2)\!\setminus\!\{1\} \\
      \Expect{\!\log\!\left(\!\frac{1}{1\!+\!Y_{ub}}\!\right)\!}         &\geq \log\!\left(\!\frac{1}{1\!+\!\Expect{Y_{ub}}}\!\right),          & \alpha& =1 \\
    \Expect{1+Y_{ub}}               &=    1+\Expect{Y_{ub}},                & \alpha& =2 \\
    \Expect{\!\frac{(1\!+\!Y_{ub})^{\alpha\!-\!1}}{1\!-\!\alpha}\!}  &\leq \frac{(1\!+\!\Expect{Y_{ub}})^{\alpha\!-\!1}}{1\!-\!\alpha},   & \alpha& >2.
  \end{align}

  The above inequalities may then be employed similarly to the remaining \(\alpha\) (omitted for brevity), yielding the following bounds on the expected value of the random assignment \(\Zbf\):
  \begin{equation}\label{eq:num-ssa-expect}
    \Expect{\fRSO(\Zbf)} \begin{cases}
      \geq \fRSO(\xbf^*), & \alpha \in [0,2) \\
      = \fRSO(\xbf^*), & \alpha = 2 \\
      \leq \fRSO(\xbf^*), & \alpha > 2
    \end{cases}.
  \end{equation}

  Now, we note that the expectations in \eqref{num-ssa-expect} are a weighted sum of objective function values associated with rounded (integer) solutions where the weights are determined by the probability of rounding to each integer solution \(\zbf\).
  When \(\alpha\in[0,2]\), we use this fact to show that any \(\zbf\) with positive rounding probability must also be an optimal solution: \(\fRSO(\zbf) = \fRSO(\xbf^*)\).
  Let \(\zbf\) be any integer solution with positive rounding probability.
  If \(\fRSO(\zbf) > \fRSO(\xbf^*)\), then we have contradicted the optimality of \(\xbf^*\); it follows that we must have \emph{i)} \(\fRSO(\zbf) \leq \fRSO(\xbf^*)\) and \emph{ii)} \(\Expect{\fRSO(\Zbf)} = \fRSO(\xbf^*)\).
  Next, we note that the event \(\fRSO(\zbf) < \fRSO(\xbf^*)\) is precluded by the previous restriction of the expectation's support and value.
  We must therefore conclude that \(\fRSO(\zbf) = \fRSO(\xbf^*)\).

  Unfortunately, this argument does not extend to \(\alpha>2\) as Jensen's inequality yields a bound in the opposite direction.
\end{IEEEproof}

\subsection{\prpref{num-rssa-optalloc-convexity} (Non-convexity of \texorpdfstring{\(\fRSO\)}{fRSO})}\label{prf:num-rssa-optalloc-convexity}

\begin{IEEEproof}
  As mentioned by Ye \ea \cite[Eq.~10]{YeRonChe2013}, \(\fRSO\) is a convex program under \(\log\)-utility (\(\alpha=1\)).

  When \(\alpha = 0\), each summand (\(\max_{u} r_{ub}x_{ub}\)) is convex, and thus the objective function, being a non-negative weighted sum of convex functions, is also convex, which makes \(\fRSO\) a convex maximization problem.


  When \(\alpha > 0\) and \(\alpha \neq 1\) we show each summand in \(\fRSO\), of the form \(g(x,y) = x(1+y)^{\alpha-1}\), is non-convex.
  The set of second order partial derivative, or Hessian \(H\), of \(g(x,y)\) are:
  \begin{equation}
    H = \begin{bmatrix}
      \partial_{xx} = 0 & \partial_{xy} = (\alpha\!-\!1)(1\!+\!y)^{\alpha-2}  \\
      \partial_{yx} = \partial_{xy} & \partial_{yy} = (\alpha\!-\!1)(\alpha\!-\!2)x(1\!+\!y)^{\alpha-3}
    \end{bmatrix},
  \end{equation}
  which has eigenvalues:
  \begin{align}
    \{\lambda_1,\lambda_2\} &= \left\{\frac{1}{2}(\alpha-1)(1+y)^{\alpha-3} \right. * \\
    &\quad \left. \left((\alpha-2)x \pm \sqrt{(\alpha-2)^2 + 4(1+y)^2} \right)\right\}.
  \end{align}

  Examining their ratio \(\lambda_1/\lambda_2\), we have:
  \begin{align}
    \frac{\lambda_1}{\lambda_2} &= \frac{1 - \sqrt{\frac{(\alpha-2)^2 + 4(1+y)^2}{(\alpha-2)^2x^2}} }{ 1 + \sqrt{\frac{(\alpha-2)^2 + 4(1+y)^2}{(\alpha-2)^2x^2}} }.
  \end{align}
  The denominator is clearly positive, while the numerator is negative:
  \begin{equation}
    1 - \sqrt{\frac{1}{x^2}\left(1 + \frac{4(1+y)^2}{(\alpha-2)^2}\right)} < 1 - \sqrt{1 + \frac{4(1+y)^2}{(\alpha-2)^2}} < 0,
  \end{equation}
  which guarantees the eigenvalues of \(H\) are of mixed signs, the indefiniteness of \(H\), and the non-convexity of \(g(x,y)\).
  In the special case of \(\alpha = 2\), the Hessian \(H\) consists of the ones in the off-diagonal with eigenvalues \(\{\lambda_1,\lambda_2\} = \{-1,1\}\).
  In the special case of \(\alpha \to \infty\), the summand also fits the form \(g(x,y) = x(1+y)^{\alpha-1}\) (with \(\alpha=0\)).
  Since \(\fRSO\) is a weighted summation of non-convex functions, we conclude that \(\fRSO\) is in general non-convex.
\end{IEEEproof}

\subsection{\prpref{num-ssa-unialloc} (SSA NUM Problem w/ Uniform Allocation)}\label{prf:num-ssa-unialloc}

\begin{IEEEproof}
  When \(\alpha \rightarrow \infty\), \eqref{num-ssa} becomes a minimax problem:
  \begin{align}
    &\argmax_{\zbf} \lim_{\alpha\rightarrow\infty} \sum_{u} \utility{R_{u}(\ybf(\zbf))} \\
    &\quad = \argmin_{\zbf} \lim_{\alpha\rightarrow\infty} \frac{1}{\alpha-1} \sum_{u} \left(\sum_{b} r_{ub}y_{ub}(\zbf)\right)^{\alpha-1} \\
    &\quad \stackrel{(a)}{=} \argmin_{\zbf} \max_{u} \sum_{b} r_{ub}z_{ub} \left(1 + \sum_{v}^{v\neq u} z_{vb} \right)^{-1} \\
    &\quad = \argmax_{\zbf} \min_{u} \left(\sum_{b} r_{ub}z_{ub} \left(1 + \sum_{v}^{v\neq u} z_{vb} \right)^{-1}\right)^{-1},
  \end{align}
  where in (a) we apply the uniform allocation policy for \(\ybf(\zbf)\) given in \prpref{num-ssa-solution} before taking the limit.

  For finite \(\alpha > 0\), we have:
  \begin{align}
    \sum_{u} \utility{ R_{u}(\ybf(\zbf)) }
    &= \sum_{u} \utility{ \sum_{b} r_{ub}y_{ub}(\zbf) } \\
    &= \sum_{ub} z_{ub} \utility{ r_{ub}y_{ub}(\zbf) },
  \end{align}
  where we move the summation over \(b\) outside of the utility function as \(y_{ub}(\zbf)\) is active for exactly one summand.

  Finally, when \(\alpha = 0\), we continue the above:
  \begin{align}
    \sum_{u} \utility{ R_{u}(\ybf(\zbf)) }
    &= \sum_{ub} r_{ub}y_{ub}(\zbf)z_{ub} \\
    &= \sum_{ub} r_{ub}z_{ub}\left(1 + \sum_{v}^{v\neq u} z_{vb}\right)^{-1},
  \end{align}
  noting that the optimal solution \(\ybf(\zbf)\) from \prpref{num-ssa-solution} has each BS \(b\) allocate all resources to associated users with the highest rate.
\end{IEEEproof}

\subsection{\thmref{num-ssa-unialloc-int-gap} (Integrality Gap of \texorpdfstring{\(\fSU\)}{fSU})}\label{prf:num-ssa-unialloc-int-gap}

\begin{IEEEproof}
  Let \(\xbf^*\) be an optimal solution to \(\fRSSA\).
  Consider the following randomized rounding scheme from \(\Xmc\) to \(\Zmc\).
  Let \(\Xbf = (X_u, u\in\Umc)\) represent a random feasible SSA solution where each \(X_u \in \{1,\dots,|\Bmc|\}\) indicates the index of the BS to which MU \(u\) is assigned.
  Let each \(X_u\) be independently chosen via a distribution induced by the optimal solution \(\xbf^*\): \(\Prob{X_u = b} = x_{ub}^*\).
  Next, let \(\Zbf = (Z_{ub}, u\in\Umc,b\in\Bmc)\) be indicator \rv's formed from \(\Xbf\): \(Z_{ub} = \mathbf{1}_{\{X_u = b\}}\), and let \(Y_{ub} = \sum_{v}^{v\neq u} Z_{vb}\).
  We now bound the expected value of \(\fRSSA(\Zbf)\).

  For the case of \(\alpha=1\), we carry out the steps explicitly:
  \begin{align}
    &\Expect{f_{\alpha=1}^{\textup{RSSA}}(\Zbf)} = \Expect{\sum_{ub} Z_{ub} \left(\log\!\left( r_{ub} \right) - \log\!\left(1 \!+\! Y_{ub}\right) \right)} \\
    &\qquad \stackrel{(a)}{=} \sum_{ub} \Expect{Z_{ub}} \left(\log\!\left( r_{ub} \right) - \Expect{\log\!\left(1 + Y_{ub}\right)} \right) \\
    &\qquad \stackrel{(b)}{\geq} \sum_{ub} \Expect{Z_{ub}} \left(\log\!\left( r_{ub} \right) - \log\!\left(1 + \Expect{Y_{ub}}\right) \right) \\
    &\qquad \stackrel{(c)}{=} \sum_{ub} x_{ub}^* \left(\log\!\left( r_{ub} \right) - \log\!\left(1 + \sum_{v}^{v\neq u} x_{vb}^*\right) \right) \\
    &\qquad = f_{\alpha=1}^{\textup{RSSA}}(\xbf^*),
  \end{align}
  where we have used
  \emph{(a)} the independence of \(Z_{ub}\) and \(Y_{ub}\),
  \emph{(b)} Jensen's inequality and the concavity of \(\log(1+Y_{ub})\), and
  \emph{(c)} \(\Expect{Y_{ub}} = \sum_{v}^{v\neq u} \Expect{X_{vb}}\).

  To generalize this argument over \(\alpha\), we define \(g_{\alpha}(y)\):
  \begin{align}
    g_{\alpha}(y) \equiv \begin{cases}
      \frac{1}{1-\alpha} y^{\alpha-1},    & \alpha \neq 1 \\
      \log(1/y),                          & \alpha = 1
    \end{cases},
  \end{align}
  and note that over \(y \in \Rbb_+\), it is convex for \(\alpha \in [0,2)\), linear for \(\alpha=2\), and concave for \(\alpha > 2\).
  %
  %
  Note that the composition of \(g(y)\) with the affine mapping \(y = 1 + Y_{ub} = 1+\sum_{v}^{v\neq u} Z_{vb}\) preserves the concavity/convexity of \(g\) \cite[\S{}3.2.2]{BoyVan2004}.
  Using Jensen's inequality on \(g_{\alpha}(y)\) and expanding \(y\) we obtain the following inequalities:
  \begin{align}
    \Expect{\!\frac{(1\!+\!Y_{ub})^{\alpha\!-\!1}}{1\!-\!\alpha}\!}  &\geq \frac{(1\!+\!\Expect{Y_{ub}})^{\alpha\!-\!1}}{1\!-\!\alpha},   & \alpha& \in [0,2)\!\setminus\!\{1\} \\
      \Expect{\!\log\!\left(\!\frac{1}{1\!+\!Y_{ub}}\!\right)\!}         &\geq \log\!\left(\!\frac{1}{1\!+\!\Expect{Y_{ub}}}\!\right),          & \alpha& =1 \\
    \Expect{1+Y_{ub}}               &=    1+\Expect{Y_{ub}},                & \alpha& =2 \\
    \Expect{\!\frac{(1\!+\!Y_{ub})^{\alpha\!-\!1}}{1\!-\!\alpha}\!}  &\leq \frac{(1\!+\!\Expect{Y_{ub}})^{\alpha\!-\!1}}{1\!-\!\alpha},   & \alpha& >2.
  \end{align}

  The above inequalities may then be employed similarly to the remaining \(\alpha\) (omitted for brevity), yielding the following bounds on the expected value of the random assignment \(\Zbf\):
  \begin{equation}\label{eq:expect}
    \Expect{\fRSSA(\Zbf)} \begin{cases}
      \geq \fRSSA(\xbf^*), & \alpha \in [0,2) \\
      = \fRSSA(\xbf^*), & \alpha = 2 \\
      \leq \fRSSA(\xbf^*), & \alpha > 2
    \end{cases}.
  \end{equation}

  Now, we note that the expectations in \eqref{expect} are a weighted sum of objective function values associated with rounded (integer) solutions where the weights are determined by the probability of rounding to each integer solution \(\zbf\).
  When \(\alpha\in[0,2]\), we use this fact to show that any \(\zbf\) with positive rounding probability must also be an optimal solution: \(\fRSSA(\zbf) = \fRSSA(\xbf^*)\).
  Let \(\zbf\) be any integer solution with positive rounding probability.
  If \(\fRSSA(\zbf) > \fRSSA(\xbf^*)\), then we have contradicted the optimality of \(\xbf^*\); it follows that we must have \emph{i)} \(\fRSSA(\zbf) \leq \fRSSA(\xbf^*)\) and \emph{ii)} \(\Expect{\fRSSA(\Zbf)} = \fRSSA(\xbf^*)\).
  Next, we note that the event \(\fRSSA(\zbf) < \fRSSA(\xbf^*)\) is precluded by the previous restriction of the expectation's support and value.
  We must therefore conclude that \(\fRSSA(\zbf) = \fRSSA(\xbf^*)\).

  Unfortunately, this argument does not extend to \(\alpha>2\) as Jensen's inequality yields a bound in the opposite direction.
\end{IEEEproof}